\documentclass[graybox]{svmult}

\usepackage{mathptmx}       % selects Times Roman as basic font
\usepackage{helvet}         % selects Helvetica as sans-serif font
\usepackage{courier}        % selects Courier as typewriter font
\usepackage{type1cm}        % activate if the above 3 fonts are
\usepackage{makeidx}         % allows index generation
\usepackage{graphicx}        % standard LaTeX graphics tool
\usepackage{multicol}        % used for the two-column index
\usepackage[bottom]{footmisc}% places footnotes at page bottom

\usepackage{natbib}  

\makeindex             % used for the subject index
\def\etal{et al.}

\def\teff{\ifmmode T_{\rm eff} \else $T_{\mathrm{eff}}$\fi}

\def\ltsima{$\buildrel<\over\sim$}
\def\lsim{\lower.5ex\hbox{\ltsima}}

\newcommand{\hii}{H~{\sc ii}}
\newcommand{\ha}{\ifmmode {\rm H}\alpha \else H$\alpha$\fi}
\newcommand{\hb}{\ifmmode {\rm H}\beta \else H$\beta$\fi}
\newcommand{\lya}{\ifmmode {\rm Ly}\alpha \else Ly$\alpha$\fi}

\newcommand{\heii}{He~{\sc ii}}
\newcommand{\Heiiuv}{He~{\sc ii} $\lambda$1640}
\newcommand{\Heiiopt}{He~{\sc ii} $\lambda$4686}

\newcommand{\ebv}{\ifmmode E_{\rm B-V} \else $E_{\rm B-V}$ \fi}
\def\micron{$\mu$m}
\def\kms{km s$^{-1}$}

\def\msun{\ifmmode M_{\odot} \else M$_{\odot}$\fi}
\def\msunyr{\ifmmode M_{\odot} {\rm yr}^{-1} \else M$_{\odot}$ yr$^{-1}$\fi}
\def\zsun{\ifmmode Z_{\odot} \else Z$_{\odot}$\fi}
\def\lsun{\ifmmode L_{\odot} \else L$_{\odot}$\fi}

\def\mup{\ifmmode M_{\rm up} \else M$_{\rm up}$\fi}
\def\mlow{\ifmmode M_{\rm low} \else M$_{\rm low}$\fi}
\newcommand{\oh}{\ifmmode 12 + \log({\rm O/H}) \else$12 + \log({\rm
O/H})$\fi}
\newcommand{\nii}{[N~{\sc ii}]}

\newcommand{\oiii}{[O~{\sc iii}]}
\def\Oiii{[O~{\sc iii}] $\lambda$5007}
\def\Oiiiab{[O~{\sc iii}] $\lambda\lambda$4959,5007}
\def\flyf{\ifmmode f_{\rm Lyf} \else $f_{\rm Lyf}$\fi}
\def\pz{\ifmmode P(z) \else $P(z)$\fi}
\def\ki2{\ifmmode \chi^2 \else $\chi^2$\fi}
\def\zphot{\ifmmode z_{\rm phot} \else $z_{\rm phot}$\fi}

\newcommand{\xphot}{\ifmmode x_\gamma \else $v_\gamma$\fi}
\newcommand{\xobs}{\ifmmode x_{\rm obs} \else $x_{\rm obs}$\fi}
\newcommand{\xcmf}{\ifmmode x_{\rm CMF} \else $x_{\rm CMF}$\fi}
\newcommand{\vexp}{\ifmmode V_{\rm exp} \else $V_{\rm exp}$\fi}
\newcommand{\vmax}{\ifmmode V_{\rm max} \else $V_{\rm max}$\fi}
\newcommand{\nh}{\ifmmode N_{\rm HI} \else $N_{\rm HI}$\fi}
\newcommand{\dv}{\ifmmode \Delta v({\rm em-abs}) \else $\Delta v({\rm em}-{\rm abs})$\fi}

\def\fesc{\ifmmode f_{\rm esc} \else $f_{\rm esc}$\fi}

\def\frellya{\ifmmode f^{\rm rel}_{\rm{Ly}\alpha} \else $f^{\rm rel}_{\rm{Ly}\alpha}$\fi}
\def\wlya{$W_{\rm{Ly}\alpha}$}

\newcommand{\qh}{\ifmmode q({\rm H}) \else $q({\rm H})$\fi}
\newcommand{\qhe}{\ifmmode q({\rm He^0}) \else $q({\rm He^0})$\fi}
\newcommand{\qhep}{\ifmmode q({\rm He^+}) \else $q({\rm He^+})$\fi}
\newcommand{\Qh}{\ifmmode Q({\rm H}) \else $Q({\rm H})$\fi}
\newcommand{\Qhe}{\ifmmode Q({\rm He^0}) \else $Q({\rm He^0})$\fi}
\newcommand{\Qhep}{\ifmmode Q({\rm He^+}) \else $Q({\rm He^+})$\fi}
\newcommand{\Qhtwo}{\ifmmode Q({\rm LW}) \else $Q({\rm LW})$\fi}
\newcommand{\qrathe}{\ifmmode q({\rm He^0})/q({\rm H}) \else $q({\rm He^0})/q({\rm H})$\fi}
\newcommand{\qrathep}{\ifmmode q({\rm He^+})/q({\rm H}) \else $q({\rm He^+})/q({\rm H})$\fi}
\newcommand{\Qrathe}{\ifmmode Q({\rm He^0})/Q({\rm H}) \else $Q({\rm He^0})/Q({\rm H})$\fi}
\newcommand{\Qrathep}{\ifmmode Q({\rm He^+})/Q({\rm H}) \else $Q({\rm He^+})/Q({\rm H})$\fi}
\newcommand{\tbb}{\ifmmode T_{\rm bb} \else $T_{\rm bb}$ \fi}
\newcommand{\Qhave}{\ifmmode \overline{Q}({\rm H}) \else $\overline{Q}({\rm H})$\fi}
\newcommand{\Qheave}{\ifmmode \overline{Q}({\rm He^0}) \else $\overline{Q}({\rm He^0})$\fi}
\newcommand{\Qhepave}{\ifmmode \overline{Q}({\rm He^+}) \else $\overline{Q}({\rm He^+})$\fi}
\newcommand{\Qhtwoave}{\ifmmode \overline{Q}({\rm H}_2) \else $\overline{Q}({\rm H}_2)$\fi}
\newcommand{\Qratheave}{\ifmmode \overline{Q}({\rm He^0})/\overline{Q}({\rm H}) \else $\overline{Q}({\rm He^0})/\\
overline{Q}({\rm H})$\fi}
\newcommand{\Qrathepave}{\ifmmode \overline{Q}({\rm He^+})/\overline{Q}({\rm H}) \else $\overline{Q}({\rm He^+})/\
\overline{Q}({\rm H})$\fi}
\newcommand{\zcoll}{\ifmmode Z_{\rm coll} \else $Z_{\rm coll}$\fi}

\def\mup{\ifmmode M_{\rm up} \else M$_{\rm up}$\fi}
\def\mlow{\ifmmode M_{\rm low} \else M$_{\rm low}$\fi}

\def\ewlya{$EW({\rm{Ly}\alpha})$}

\def\hii{H{\sc ii}}
\def\oiii{O{\sc iii}}
\def\nii{N{\sc ii}}

\newcommand{\la}{\raisebox{-0.5ex}{$\,\stackrel{<}{\scriptstyle\sim}\,$}}
\newcommand{\ga}{\raisebox{-0.5ex}{$\,\stackrel{>}{\scriptstyle\sim}\,$}}

\begin{document}

%\title*{Predicting observables of the first galaxies: from synthesis models to observables}
% For Springer:
%\title*{Evolutionary synthesis models as a tool and guide towards the first galaxies}
% For astro-ph:
\title*{Evolutionary synthesis models as a tool and guide towards the first galaxies$^\star$}
\titlerunning{Evolutionary synthesis models as a tool and guide towards the first galaxies}
\author{Daniel Schaerer}
\institute{
% For astro-ph:
$\star$  Chapter to appear in ``The First Galaxies -- Theoretical Predictions and Observational Clues", Eds.\ T.\ Wiklind, V.\ Bromm, B.\ Mobasher,
Springer Verlag \\
Daniel Schaerer \at 
Observatoire de Gen\`eve, Universit\'e de Gen\`eve, 51 Ch. des Maillettes, 1290 Versoix, Switzerland;
CNRS, IRAP, 14 Avenue E. Belin, 31400 Toulouse, France \email{daniel.schaerer@unige.ch}}
%\and Name of Second Author \at Name, Address of Institute \email{name@email.address}}
\maketitle
%%%%%%%%%%%%%%%%%%%%%%%%%%%%%%%%%%%%%%%%%%%%%%%%%%

\abstract{We summarize the principles and fundamental ingredients of evolutionary synthesis models, which are
stellar evolution, stellar atmospheres, the IMF, star-formation histories, nebular emission, and 
also attenuation from the ISM and IGM. The chapter focusses in particular on issues of 
importance for predictions of metal-poor and Population III dominated galaxies. 
\newline
We review recent predictions for the main physical properties and related observables of star-forming galaxies 
based on up-to-date inputs. The predicted metallicity dependence of these quantities
and their physical causes are discussed. The predicted observables include in particular
the restframe UV--to-optical domain with continuum emission from stars and the ionized ISM, as well
as emission lines from H, He, and metals.
\newline
Based on these predictions we summarize the main observational signatures (emission line strengths, colors etc.),
which can be used to distinguish ``normal" stellar populations from very metal-poor objects or even Pop III.
\newline
Evolutionary synthesis models provide an important and fundamental tool for studies of galaxy formation and evolution, 
from the nearby Universe back to first galaxies. They are used in many applications to interpret
existing observations, to predict and guide future missions/instruments, and to allow direct comparisons
between state-of-the-art galaxy simulations and observations.}

%%%%%%%%%%%%%%%%%%%%%%%%%%%%%%%%%%%%%%%%%%%%%%%%%%
\section{Introduction}

Evolutionary synthesis models, first pioneered  by \citet{1968ApJ...151..547T,1980FCPh....5..287T},
are a simple, but fundamental tool to predict the emission of integrated stellar populations, such 
as those of distant galaxies. Their basic immediate objective is to predict the total spectrum 
or the spectral energy distribution (hereafter SED) emitted e.g.\ by a galaxy or by another ensemble of 
stars. 

Synthesis models are generally used to interpret observations of integrated stellar populations, i.e.\
to infer their physical properties -- such as the total stellar mass, star formation rate, age, attenuation etc. --
from comparisons between models and observations. Often synthesis models are also used to 
predict/guide future observations, since from our knowledge of star formation, stellar evolution,
and atmospheres, one is able to predict a large number of observables for a very broad range
of parameters (ages, star formation histories, stellar initial mass functions, metallicities, redshifts etc.).
This is particularly true in the present context related to the first galaxies, which are at the
limit or beyond the reach of present-day facilities, and where predictions are needed to plan
future missions and devise observational strategies to search for these distant, ``exotic'' objects.

Focussing on emission from stars and the ISM in the ultraviolet, optical, and near-IR domain
(taken in their rest frame) -- the spectral range where usually stars dominate the integrated emission 
of galaxies -- we will describe and summarize the main ingredients (inputs) of synthesis models,
as well as the basic assumptions and related uncertainties (Sect.\ \ref{s_intro}).
In Sect.\ \ref{s_predict} we show how and why physical properties of stellar populations and corresponding observables 
vary with metallicity, and which main differences are expected between ``first galaxies", metal-free
(PopIII), metal-poor, and present-day stellar populations. Methods used to distinguish/select observationally
PopIII-dominated and similar objects from ``normal'' galaxies are reviewed in Sect.\ \ref{s_obs}.
Brief conclusions are given in Sect.\ \ref{s_conclude}.

%%%%%%%%%%%%%%%%%%%%%%%%%%%%%%%%%%%%%%%%%%%%%%%%%%
\section{Synthesis models -- basic ingredients and assumptions}
\label{s_intro}

Popular, widely used synthesis models include the models of Bruzual and Charlot \citep{BC03},
the {\em Starburst99} models specialized for young stellar populations/starbursts \citep{leitherer99},
the P\'EGASE models of \citet{fioc99},
and also recent models including a special treatment of TP-AGB stars 
\citep{2005MNRAS.362..799M,2006ApJ...652...85M}.
Recent reviews on synthesis models, including some basics as well as topics for current and future
improvements, have e.g.\ been presented by
%Bruzual 2003 IAC Winterschool
\citet{2003ghr..conf..185B,2011RMxAC..40...36B},
\citet{2011IAUS..277..158M}, and
\citet{2011arXiv1111.5204L}.
Papers presenting tests and confrontation of synthesis model predictions with basic
observations of young stellar populations include e.g.\ \citet{2001MNRAS.325...60C,2003ghr..conf..185B,2010MNRAS.403..780C}.

Schematically, the following needs to be known (or assumed) to be able to predict
the spectrum of integrated stellar populations:
\begin{enumerate}
\item {\bf Stellar evolution}: A description of the evolution (in time) of stars in the HR-diagram as a function of their
initial mass, metallicity (chemical composition), and other parameters which may govern their evolution
(e.g.\ initial rotation rate, magnetic field).
\item {\bf Stellar atmospheres}: A description of the emergent spectrum (over the spectral range of interest) of individual 
stars at all phases of their evolution.
\item {\bf The stellar initial mass function (IMF),} which determines the relative distribution of stars of different masses 
at the time of formation.
\item {\bf The star formation history (SFH)} of the galaxy, describing the history of the amount of stars (commonly expressed in
mass formed per unit time), i.e.\ the star formation rate (SFR) as a function of time.
\item {\bf Nebular emission}, i.e.\ the emission from \hii\ regions nearby massive star-forming regions, which -- in general --  cannot 
be separated from the stellar emission.
\item {\bf Attenuation} within the intervening interstellar medium (ISM) of the galaxy. For a simple prescription this
implies that we need to know the attenuation law (i.e.\ it's dependence on wavelength), and the amount of attenuation
at a given reference wavelength.
\item {\bf Intergalactic medium opacity:} Finally, since the photons emitted by a distant galaxy also travel through the
intergalactic medium (IGM), its transmission properties must be known/specificed.
\end{enumerate}

{\em Items 1--3} describe the properties of stars and ensembles thereof. So-called simple stellar populations (SSPs),
corresponding to an ensemble of stars formed at the same time, represent the basic units. Predictions
for SSPs are widely distributed in the literature. In practice SSPs
may represent stellar populations of stellar clusters where the age spread between stars is small.
For any arbitrary, given star formation history (4), the  predicted spectrum can be derived from SSPs
by convolution.
Nebular emission (5) is important to properly describe star-forming galaxies, where the contribution
from young massive stars is significant. 
The remaining items (6, 7)  describe the way the emitted spectrum is altered both at the galaxy scale
and through the IGM on its way to the observer.

Evolutionary synthesis models traditionally describe the emission from stars 
(plus emission from surrounding \hii\ regions in some extensions). By construction such models are thus
usually tailored to the (rest-frame) UV--optical--near-IR part of the electromagnetic spectrum, where emission
from stars (+nebulae)  dominate. This is the domain on which the present text is focussed.
Extensions of these models to other wavelength domains, e.g.\ to X-rays or to the IR--radio, have also been 
constructed by \citet{1991A&AS...88..399M,2002A&A...392...19C}.
Since focussed on primeval galaxies and on their observability, we will here emphasize in 
particular very low metallicities (necessarily more important in the early Universe) and relatively massive stars,
which dominate the rest-frame UV emission in strongly star-forming galaxies.

Let us now briefly discuss each of these ``ingredients'' of evolutionary synthesis models with a special
emphasis on first/distant galaxies.

\begin{figure}[htb]
\sidecaption
\includegraphics[scale=.4]{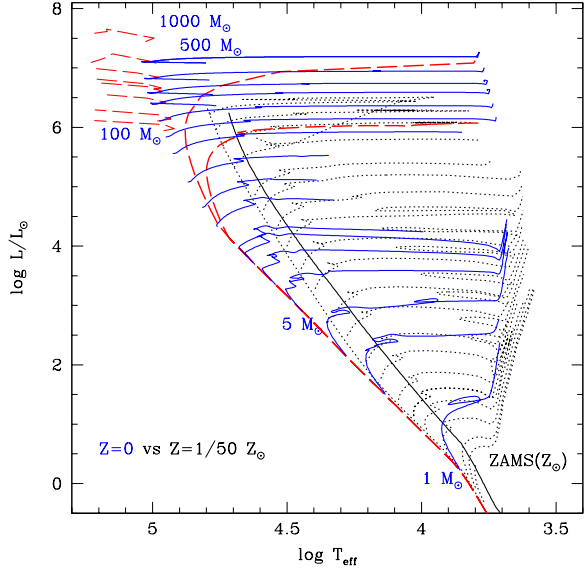}
\caption{HR--diagram for metal free ($Z=0$, solid and long-dashed lines) 
and low metallicity ($Z=1/50 \zsun$, dotted) stars.
Isochrones of 2 and 4 Myr for $Z=0$ tracks without mass loss are also 
plotted (long-dashed). The short dashed high mass tracks evolving blueward
of the ZAMS are computed assuming strong mass loss.
The position of the ZAMS at solar metallicity (\zsun) from Schaller \etal\ (1992) is 
shown by the solid line.
%Pop III tracks and isochrones from Marigo \etal\ (2001, no mass loss) and 
%Klapp (1983), and El Eid \etal\ (1983, strong mass loss).
%$Z=0.0004$ tracks for 0.8 to 150 \msun\ from Lejeune \& Schaerer (2001).
Other tracks and isochrones are also shown \citep[see legend in][]{Scha02}.
Note the important shift of the ZAMS to high \teff\ from low metallicity
to $Z=0$, as well as the rapid redward evolution of the massive stars.
From \citet{Scha02}.
}\label{fig_hrd}
\end{figure}

% % % % % % % % % % % % % % % % % % % % % % % % % % % % % % % 
\subsection{Stellar evolution}
Stellar evolution models have seen a ``boom'' in the 1990ies, when new radiative opacities were published,
triggering extensive evolutionary track calculations for a wide range of stellar masses and metallicities.
The most widely used of these models are the Geneva and Padova tracks
\citep[see][]{1992A&AS...96..269S,1994A&AS..103...97M,1994A&AS..106..275B}, which have
been extensively used since then for evolutionary synthesis models and other applications.
An illustration of such tracks for zero metallicity, and a comparison with 1/50 \zsun\ and solar metallicity
(\zsun) is shown in Fig.\ \ref{fig_hrd}.
One of the main, if not the main property	distinguishing PopIII stars from others is clearly apparent
from this figure: the fact that massive ($M \ga 5$ \msun) stars are much hotter than their counterparts
at non-zero metallicity. For the most massive stars their effective temperature can reach up to $\sim 10^5$ K
on the zero-age main sequence (ZAMS). These differences has several important observational consequences,
which are discussed below. 
A detailed discussion of the peculiarities of interior models of PopIII stars and the literature before 2001
is given in \citet{2001A&A...371..152M}.

Since the stellar evolution models computed in the 1990ies including numerous physical processes
among which in particular mass loss, large efforts have been undertaken to describe other processes
affecting the interior evolution of stars, such as stellar rotation, magnetic fields, and the various
transport mechanisms related to it. For reviews on these issues see e.g.\ 
\citet{2000ARA&A..38..143M,2012RvMP...84...25M}
The impact of other, maybe more ``exotic" phenomena, like possible variations of the fine-structure constant 
on stellar properties at very low metallicity have e.g.\ been explored by \citet{2010A&A...514A..62E}.

Several arguments, both observational and theoretical, indicate that the importance of rotation becomes
stronger for low metallicities \citep[see e.g.][]{1999A&A...346..459M,2006A&A...449L..27C}.
However, the predicted stellar tracks depend strongly on the initial rotational velocity (a free parameter) 
and its history (which is predicted from the model following also the evolution of angular momentum), which
are very uncertain, and difficult to constrain at (very) low metallicity. In consequence, the impact
of stellar rotation on predictions for integrated stellar populations remains currently poorly known.
Preliminary explorations of the impact of stellar rotation, mostly at solar metallicity, have been presented
by \citet{2007ApJ...663..995V}, \citet{2011arXiv1111.5204L}, and Levesque et al.\ (2012, submitted).

Recent PopIII tracks including the effects of stellar rotation and magnetic fields are presented in 
%Eckstrom 2008, Yoon et al. 2012
\citet{2008A&A...489..685E} and \citet{2012arXiv1201.2364Y}. For the initial rotational velocities
chosen in the former paper, rotation has a small impact on the evolution in the HR-diagram 
and hence on predictions from synthesis models. However, the surface abundances of these stars
are significantly modified, with several implications on  nucleosynthesis and chemical evolution.
Effects with potential impact on evolutionary synthesis models can be found for very high rotation rates
\citep[cf.][]{2012arXiv1201.2364Y},
when stars are very strongly mixed, following a nearly homogeneous evolution, which implies
much hotter temperatures and a blue-ward evolution. In this case properties such as their UV and ionizing
flux are significantly altered, leaving imprints on the  spectra predicted from stellar populations
containing such stars. However, the distribution of rotational velocities remains poorly know,
especially at very low metallicity, and hence the proportion of stars with properties significantly altered
by rotation are not known. 
\citet{Scha03} has explored in a simple way the possible impact of very hot, homogeneous
PopIII stars on the hardness of the ionizing spectrum of stellar populations.
It is quite evident that we currently do not  have a clear view of the impact stellar rotation and magnetic fields
may have on evolutionary synthesis models, quite independently of metallicity. 
Progress in these areas is ongoing.

Other  issues of general importance for synthesis models include e.g.\  the importance of thermally-pulsating
asymptotic giant branch (TP-AGB) stars, whose contribution to the integrated light is being debated
%Marigo, Bruzual, Connor
\citep[see][and references therein]{2005MNRAS.362..799M,2011ASPC..445..391M},
and the recurrent question of the importance of binary stars 
%cf. vanbeveren, Robert+, SV98, Eldridge
\citep[cf.][]{1998NewA....3..443V,SV98,2003A&A...400..429B,2006ApJ...641..252D,2011arXiv1106.4311E}
For our present objective, first galaxies, uncertainties related to TP-AGB stars are not a major concern,
since we are dealing with young stellar populations and focussing on the UV to optical (rest-frame) domain,
where these stars can safely be neglected. 
The evolution of close massive binary stars can alter the predicted spectra of young stellar populations,
as illustrated e.g.\ recently by \citet{2011arXiv1106.4311E}.
However, the physics of these stars is even more complex than that of single stars, and depends on 
a number of additional, poorly constrained parameters. 
%
%dsNEW
In any case, it is important that the predictions from stellar evolution models be compared
to and tested against direct stellar observations in our Galaxy and in the nearby Universe.
This includes e.g.\ observations of individual stars in clusters, their basic stellar parameters and
surface abundances, color-magnitude diagrams, integrated colors of clusters, statistics
of stars of different types and trends with metallicity etc. Such comparisons have e.g.\ been
carried out with the (non-rotating) Padova and Geneva models 
\citep[cf.][]{1994A&A...287..803M,2003ghr..conf..185B,BC03,2003ARA&A..41...15M,2005A&A...429..581M},
and are being carried out for the latest tracks including stellar rotation.
They serve thus also to ``calibrate" uncertainties or unknowns in the stellar models, and hence
to place synthesis model predictions on the best ground. 

% % % % % % % % % % % % % % % % % % % % % % % % % % % % % % % 
\begin{figure}[htb]
\sidecaption
\centering{
\includegraphics[width=0.6\textwidth]{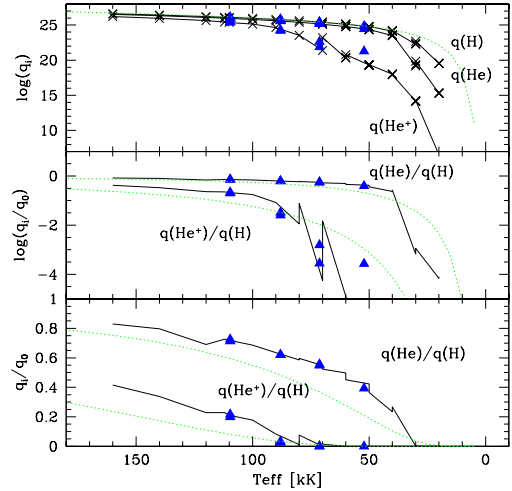}
}
\caption{Ionising photon flux per unit surface (top panel) and hardness of the ionising 
spectrum (middle and bottom panel) as a function of effective temperature 
predicted from non-LTE plane parallel {\em TLUSTY} Pop III  models 
of various \teff\ and $\log g$ (solid black lines). Triangles show calculations from the spherically 
expanding {\em CMFGEN} models for Pop III stars. 
%For comparison the pure H and He WR models of Schmutz \etal\ (1992) are also 
%shown (open squares and circles).
Green dashed lines show predictions from blackbody spectra.
Similar to Fig.\ 2 from \citet{Scha02}.}
\label{fig_qi_teff}
\end{figure}

\subsection{Stellar atmospheres}

It is well known that strong departures from local
thermodynamic equilibrium (LTE) occur in the atmospheres of hot stars. To properly predict the emergent
spectra of massive stars it is therefore essential to use non-LTE model atmospheres. 
This statement holds in particular also for low- and zero metallicities, as departures from
LTE significantly alter the level populations of H and He, which -- together with electron scattering -- 
are the main opacity sources in stars of such composition.
While generally stellar winds also affect the predicted spectra of massive star
\citep[e.g.][]{1989A&A...226..162G},
%e.g. Gabler 1989,
the mass outflow from very low metallicity stars is very low, and its effect can hence be
safely neglected to compute the observable properties of PopIII and similar massive stars
%cf. Schaerer2002, Kudritzki, KubatKritka.
\citep[cf.][]{Scha02,2002ApJ...577..389K,2006A&A...446.1039K}.
Using non-LTE plane parallel model atmospheres is therefore sufficient for this purpose.
For less extreme metallicities, appropriate model atmospheres describing main sequence stars,
Wolf-Rayet stars with strong winds, and also cooler stars must be combined 
\citep[cf.][]{2008flhs.book.....C} to achieve the most reliable synthesis model predictions
\citep[e.g.][]{leitherer99,2002MNRAS.337.1309S}.

\begin{figure}[tbh]
\sidecaption
%\centerline{\psfig{figure=15236fg20.ps,angle=90,width=10cm}}
\includegraphics[width=0.8\textwidth,angle=90,trim=0cm 13cm 1cm 1cm,clip]{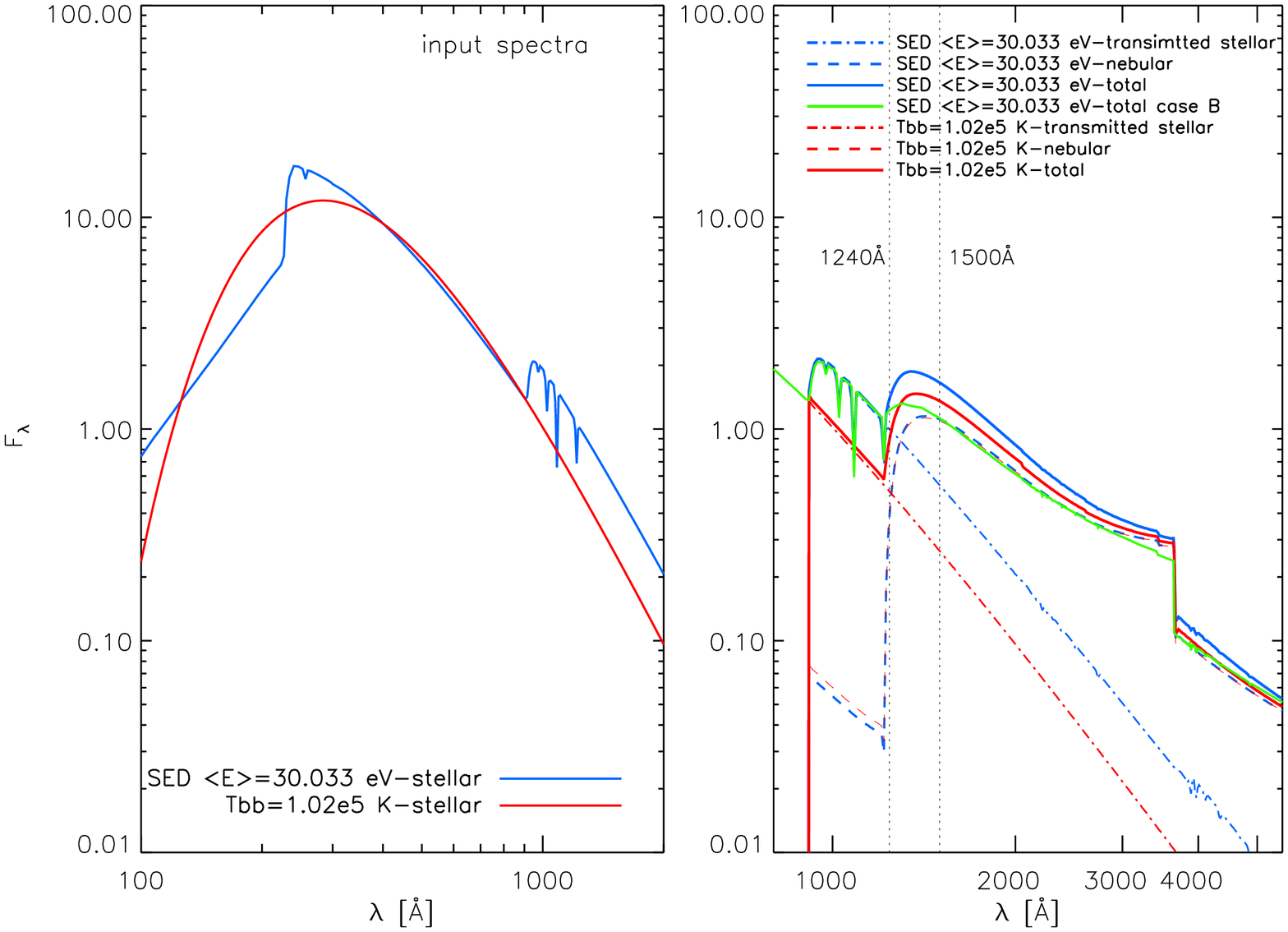}
\caption{Comparison of input stellar spectra for photoionization models showing a realistic SED 
(blue line; Pop~III, Salpeter IMF 1-100 M$_{\odot}$ at zero age)
and a black body spectrum (red line) with the same average photon energy 
($\overline E=30.033$ eV) in the Lyman continuum.
The SED model is normalized at 1240 \AA\  and
rescaled to match the same ionizing flux $Q(H)$.
Despite having the same Lyman continuum flux and average mean energy, the black body differs 
significantly, e.g.\ in the He$^+$ continuum ($\lambda < 228$ \AA), and in the observable UV.
Figure from \citet{2010A&A...523A..64R}.
}
%{\bf Right:} Input stellar spectra (dash-dotted, same colours as left panel) 
%and predicted nebular and total continua
%from our CLOUDY models for $\log(n(\rm H)) = 1$ \cmc\ and  $\log(U) = -1$ (solid lines, using the same colours as in left panel;
%cf.\ also inset for symbols). 
%The green solid line shows the total (stellar + nebular) SED from our
%evolutionary synthesis model assuming constant nebular density and temperature and 
%case~B. Notice the difference between case~B and the CLOUDY result due to enhanced 2$\gamma$ emission.}
\label{cont_comp}
\end{figure}

The predicted ionizing fluxes of H, He, and He$^+$ from PopIII stars as a function of effective temperature
is shown in Fig.\ \ref{fig_qi_teff}. For comparison, the same values predicted using simple black-body
spectra are also shown. While for $\teff \ga$ 40 kK black bodies provide a good approximation to the
total number of H ionising photons (i.e.\ at energies $>$ 13.6 eV), this is not the case for the shape
of the spectra at higher energies, for the number of He and He$^+$ ionizing photons, and for other
features such as the Lyman break \citep[cf.][]{Scha02}.
Also, using black-body spectra is a bad approximation to compute observable properties such
as UV fluxes and emission line strengths (equivalent widths), since black-bodies do not reproduce
the main bound-free edges, which significantly shape true stellar spectra. See Fig.\ \ref{cont_comp}
for an illustration.
In short, appropriate state-of-the-art stellar atmosphere models should be used for
reliable predictions both of individual stars and integrated stellar populations.

% % % % % % % % % % % % % % % % % % % % % % % % % % % % % % % 
\subsection{IMF}
As for other ingredients summarized above, the stellar initial mass function (IMF) is a separate topic, 
to which many papers and conferences are devoted
%e.g. ASP Series IMF, ARAA Bastian et al. 2010
\citep[see e.g.][]{1998ASPC..142.....G,2010ARA&A..48..339B,2011ASPC..440.....T}

In the nearby Universe the IMF is well described by a log-normal function for masses below 
$\sim 1$ \msun, and by a power-law above that \citep{2003PASP..115..763C},
as already found e.g.\ in the pioneering study of \citet{1955ApJ...121..161S}.
Other authors approximate the IMF by piecewise power laws \citep{2001MNRAS.322..231K}.
In our Galaxy and in other nearby systems it is found that the IMF is only weakly dependent
on environnent \citep{2003PASP..115..763C,2010ARA&A..48..339B}.
For obvious reasons it is not possible to constrain the IMF in the same manner (e.g.\
by direct star counts) outside the local Group. Hence our empirical knowledge on the IMF
and possible variations with metallicity, stellar density, UV radiation field and other factor
which may influence it, is very limited.

%For massive stars ($\ga 1 \msun$) the IMF is approximately a power law with slope close to 
%-2.3, the value determined already by Salpeter.

For many/most observables measurable for high-redshift, star-forming galaxies only
the upper part of the IMF is relevant, since massive stars largely dominate the
luminosity in integrated stellar populations. As a reasonable rule of thumb one may therefore
simply rescale the results derived for one given IMF to those expected for another IMF 
differing in the domain of low stellar masses (i.e.\ typically at $M \ll 5$ \msun).  This is e.g.\
used to ``correct" results from synthesis models computed for a Salpeter IMF extrapolated
down to 0.1 \msun\ to those expected for a more realistic IMF below \la\ 1 \msun.
However, the applicability of this ``rule" depends on the age and star-formation history, and 
on the observable used.

Several studies, following different arguments have suggested deviations from the ``local" IMF
in distant galaxies with deviations depending on environment, galaxy type, gas density, metallicity
and other factors.
For an overview over this vast topic the reader is referred to a recent conference
proceedings on the IMF \citep{2011ASPC..440.....T}.
For metal-poor environments and Population III stars, simulations have long suggested
a preferential mass scale much higher than in the local Universe, with "typical" masses
of the order of $\sim$ 10--100 \msun.
See the chapters of Glover and Johnson in this book, and also recent conference 
proceedings on the first stars \citep[e.g.][]{2008IAUS..255.....H,2010AIPC.1294.....W}.
This difference, basically due to significantly reduced cooling of metal-poor or free
gas, is predicted to occur below a certain critical metallicity $Z_{\rm crit} \le 10^{-5\pm1}$ \zsun\
\citep{schne02,schne03}. If correct, this implies a different, more massive IMF,
at least below this metallicity threshold. Accordingly various parametrisations of the IMF have been
used in this domain \citep[see e.g.][]{Larson98,Tumlinson06,2010A&A...523A..64R} .
However, as already mentioned above, our knowledge of the IMF, especially in such extreme
conditions remains very limited, and synthesis models can simply assume different cases 
and examine their implications.

Synthesis models also generally make the assumption of a continuous, well-populated IMF,
which is correct for stellar systems with a large enough number of stars. However, 
in ``small" stellar populations the analytic statistical description of the IMF will break down, 
and sampling of the IMF with a small/finite number of stars may lead to 
significant differences, due to the progressive absence of massive stars. Examples
of such  ``stochastic" IMF effects on colors, ionizing fluxes and others have been illustrated by 
various authors
\cite[see][and references therein]{2000A&A...360L...5C,2006A&A...451..475C,2002IAUS..207..616B}.
Typically stochastic effects are found to be significant for cluster masses 
below \la $10^5$ \msun\ or for low star formation rates SFR \la\ 1 \msunyr, 
although this limit again depends on the observable of interest \citep{2000A&A...360L...5C,2011ApJ...741L..26F}.
For small entities, such as the first galaxies, stochasticity may thus be important for some cases.

\begin{figure}[htb]
\sidecaption
\centering{
\includegraphics[width=0.6\textwidth]{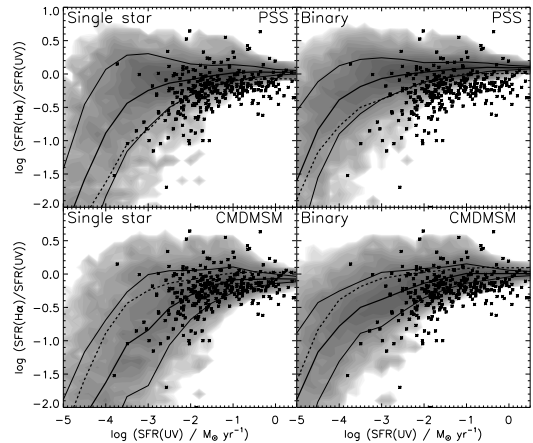}
}
\caption{The ratio of SFR measured by \ha\ and UV fluxes versus the SFR from UV flux. The asterisks are the observations of \citet{2009ApJ...706..599L},
while the shaded region show the density of our individual realisations of synthetic galaxies. The thick solid lines indicate the mean ratios for the synthetic galaxies and their $1\sigma$ limits. The dashed lines show the mean ratios for the other IMF filling method with the same stellar population. The upper and lower panels distinguish a stochastic sampling of the IMF and a sampling 
including a description of the cluster mass function.
Left/right panels distinguish synthesis model including single/(single+binary) stars.
%panels are for PSS and the lower panels are for CMDMSM. While the left panels are for a single star population and the right panels are for binary populations
%
From \citet{2011arXiv1106.4311E}.}
\label{fig_eldridge}
\end{figure}

The question of a stochastically sampled IMF is also related to the concept of the
integrated galactic initial mass function (IGIMF)
introduced by Kroupa and collaborators
\citep[see e.g.][]{2007ApJ...671.1550P,2006MNRAS.365.1333W}.
Various applications of synthesis models implementing a stochastic IMF can e.g.\ be found
in \citet{2002A&A...394..443P}, % WR galaxies Pindao et al.
\citet{2011ApJ...741L..26F}, and \citet{2011arXiv1106.4311E}.
For example, a stochastic IMF leads to a significant scatter in the relative \ha\ and UV output
for populations with a low SFR (i.e.\ forming a small number of stars), as illustrated 
in Fig.\ \ref{fig_eldridge}. This is due to the fact that these two emissions originate from stars of somewhat
different mass regimes,  \ha\ being due to more massive stars than the UV continuum.
Since stochastic effects are of increasing importance for low mass / low SFR galaxies, this issue should 
be relevant  for studies of the first galaxies. The implications remain largely to be worked out.

% % % % % % % % % % % % % % % % % % % % % % % % % % % % % % % 
\subsection{Star formation history}
Star formation histories are a key ingredient for evolutionary synthesis models.
As already mentioned above,  simple stellar populations (SSPs) represent the basic units, and 
for any arbitrary star formation history (SFH), the integrated spectrum can be derived from SSPs
by convolution. Historically, simple parametrisations of the SFH have
been used, the most common one being a family of exponentially decreasing SFHs with
SFR$(t) \propto \exp(-t/\tau)$, where $\tau>0$ is a characteristic timescale.
It is well known that such star formation histories are able of reproducing the observed spectro-photometric
properties of present-day galaxies of all Hubble types
\citep{1978ApJ...219...46L,1968ApJ...151..547T,Kenn98}.

Constant SFR, corresponding to $\tau=\infty$, is a limiting case often used to
derive calibrations for the SFR from various observables (e.g.\ UV continuum,
H recombination lines, bolometric luminosity etc.). See e.g.\
\citet{Kenn98,2000bgfp.conf..389S}.

%Oti-Florales +Mas-Hesse 2011 - short and long SFR calib**

Depending on the application, other, more complex SFHs have been explored.
Many numerical simulations of galaxy formation and evolution (semi-analytical models,
hydrodynamic models and others) are now ``coupled'' with evolutionary synthesis models
to predict observable properties consistent with their (complex) star formation histories.

For distant ($z>2$) galaxies SF histories and corresponding timescales are currently vividly debated. 
For example, simulations suggest rapid growth of galaxies with increasing star formation rates
during the first Gyr of the Universe, i.e.\ at $z>6$ \citep{Finlator07,2011MNRAS.410.1703F}.
From the apparent tightness of the mass-SFR relation of galaxies at high redshift
some authors argue for rising star formation histories for galaxies down to $z \sim$ 2--3
\citep{2010MNRAS.407..830M,2011MNRAS.412.1123P}.
At somewhat lower redshift, other authors suggest e.g.\ ``delayed" star formation histories
SFR$(t) \propto t \exp(-t/\tau)$ showing both phases of increasing (for $t<\tau$) and decreasing
SFR ($t>\tau$), and suggest relatively long timescales of several 100 Myr \citep{2011ApJ...738..106W}.
%Long time scales and delayed SFHs preferred by Wuyts et al. 2011 - z~2
From the clustering of $z \sim$ 4--5 Lyman break galaxies (LBGs), \citet{2009ApJ...695..368L}
argue for relatively short duty cycles ($<0.4$ Gyr).
Even shorter timescales and declining SFHs are favored from SED analysis of Lyman break galaxies 
at $z \sim$ 3--6 with spectral templates accounting for nebular emission \citep[][de Barros et al.\ 2012, in preparation]{2011arXiv1111.6373S}.
The question of star formation histories of distant galaxies is closely related to the debate
about the main mode of star formation, i.e.\ schematically about the relative importance of (cold) accretion driven
star formation and mergers \citep[see e.g.][and references therein]{2011MNRAS.410L..42K}
In any case the question of the ``typical'' star formation history of distant galaxies is not yet settled,
and a large diversity of histories -- more complicated than simple, parametrised functions -- must occur in nature.
Furthermore it is clear that predictions from evolutionary synthesis models strongly depend
on this important quantity.

% % % % % % % % % % % % % % % % % % % % % % % % % % % % % % % 
\subsection{Nebular emission}
\label{s_nebular}
Stars (and stellar populations) described above provide a dominant, but not the sole
source of emission in the (rest-frame) UV, optical, and near-IR light of galaxies.
Emission from ionized regions (the so-called \hii\ regions) of the surrounding ISM is another
important contribution, which needs to be taken into account to describe/predict
the spectra or SEDs of galaxies in this part of the electromagnetic spectrum.
Hence this should, in general, be treated (added) to the prediction from ``standard" evolutionary synthesis
models describing stellar emission only. 

\begin{figure}[htb]
\sidecaption
\centering{
\includegraphics[width=0.6\textwidth, trim=0cm 16.85cm 8.3cm 0cm,clip]{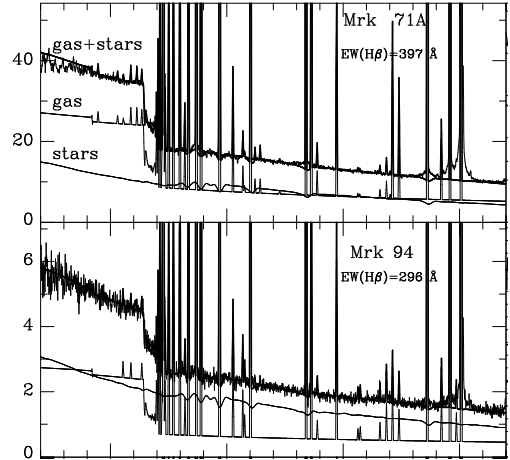}
}
\caption{Observed spectra of two metal-poor \hii\ galaxies from the sample of \citet{2006ApJ...644..890G}
focussing on the region between the UV and optical domain. Superposed are the stellar and 
gaseous (nebular continuum) contribution.
The flux is given in $F_\lambda$ units between 3200 and 5200 \AA\ with spacing of 100 \AA\ between tickmarks. 
The Balmer jump (not a ``break" here), due to nebular free-bound emission, is clearly seen around 3650 \AA; no stellar 
source is known to show such emission.
Figure adapted from \citet{2006ApJ...644..890G}.
}
\label{fig_guseva}
\end{figure}

Most galaxy types across the Hubble sequence show at least some signs of emission lines
in the optical. Their strength increases towards late types and irregular galaxies, basically due
to an increasing ratio of present over average past star formation \citep[cf.][]{Kenn98}.
Nearly by definition, nebular emission is associated with star-forming galaxies, since
-- as long as stars more massive than 5--10 \msun\ are formed -- star formation always
implies the emission of  UV photons with energies $>$ 13.6 eV capable of ionizing H and other
elements in the ISM, causing a plethora of recombination lines.
In addition several processes in \hii\ regions, including recombination and 2-photon emission from the 
2 $^2$S level of hydrogen, produce a continuum emission longward of \lya\  whose  emissivity
increases with wavelength \citep{Osterbrock06}. Hence, nebular emission implies a priori both
line and continuum emission. An illustration of the three emission components, stars, nebular lines,
and nebular continuum, observed in metal-poor, nearby galaxies is shown in Fig.\ \ref{fig_guseva}.

{\em To first order}, the strength/luminosity of both lines and nebular continuum emission depends on the 
flux/luminosity of ionizing photons in the Lyman continuum denoted here by \Qh,
as can be derived from well known nebular physics, assuming e.g.\ so-called case A or B  
and typical nebular densities and temperatures \citep{Osterbrock06}. 
In this approximation it is straightforward to predict the emission
from the major H and He recombination lines and continuum emission
for a given source spectrum, such as calculated by evolutionary synthesis models.
For other emission lines from \hii\ regions, mostly forbidden metal-lines (in the optical domain),
full photoionization models need to be computed, or other prescriptions be used.

\begin{figure}[htb]
\sidecaption
\centering{
\includegraphics[width=0.6\textwidth]{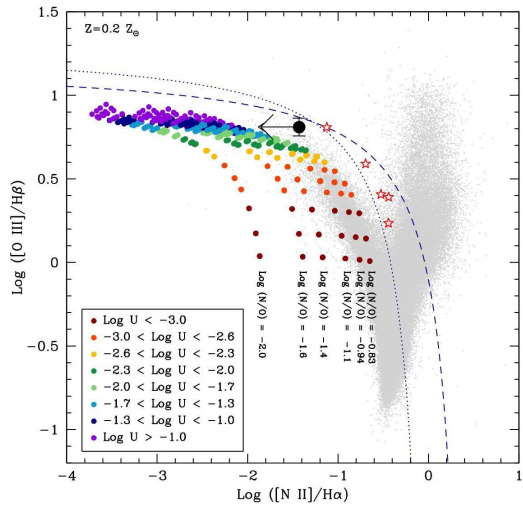}
}\caption{The [\nii] /\ha\ versus [\oiii] /\hb\  diagnostic diagram. 
Small grey points show $\sim$ 96000 objects from the Sloan Digital Sky Survey, the large black circle the peculiar,
low-metallicty $z \sim 2$ galaxy BX418 discovered by \citet{2010ApJ...719.1168E}.
Colored points are predictions from photoionization models, coded by ionization parameter as labeled at lower left. 
From \citet{2010ApJ...719.1168E}.}
\label{fig_erb}
\end{figure}

{\em To second order} nebular emission depends on the conditions in the nebula/ISM,
which are primarily described by the electron temperature and density ($T_e$, $n_e$),
and most importantly on the ionization parameter $U$. The latter depends on the radiation field,
geometry, and density, and in general the nebular conditions also depend on metallicity.
It is well known that \hii\ regions and galaxies at low metallicity show higher excitation and stronger optical 
emission lines, as illustrated e.g.\ in nebular diagnostic diagrams
\citep{1981PASP...93....5B,1987ApJS...63..295V}, an example of which is shown in Fig.\ \ref{fig_erb}.

Since both the excitation of the ionized gas and the ionizing flux increase with 
decreasing metallicity, nebular emission must on average be stronger in unevolved/early galaxies
 than in present-day galaxies. How this affects various observables will be discussed in detail below.

%TODO?? Dopita ionization parameter.

\begin{figure}[htb]
\sidecaption
\centering{
\includegraphics[width=0.6\textwidth]{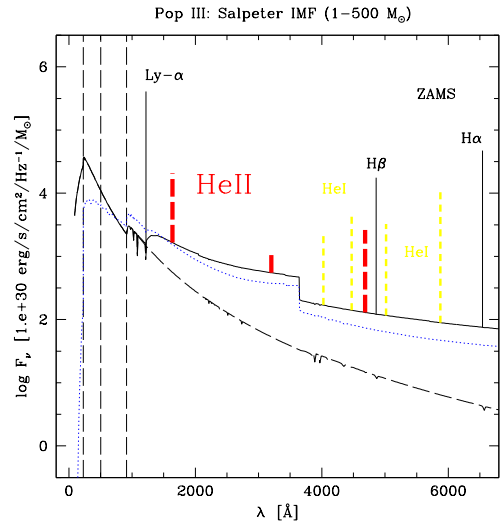}
}
\caption{Spectral energy distribution of a young (zero age) PopIII starburst with an IMF extending to 500 \msun.
The SED shown in full lines includes H and He recombination lines, nebular, and stellar continuum.
The pure stellar continuum (neglecting nebular emission) is shown by the
dashed line. For comparison the SED of the $Z=1/50 \zsun$ population
is shown by the dotted line. 
The vertical dashed lines indicate the ionisation potentials of H, He$^0$, and  He$^+$. 
Note the presence of the unique \heii\ features (shown as thick dashed lines)
and the importance of nebular continuous emission.
From \citet{Scha02}.}
\label{fig_S02}
\end{figure}

Indeed, for Population III stars and for ensembles thereof, \citet{Scha02} has demonstrated that 
nebular continuum emission  dominates the spectrum at wavelength longward of \lya,
as illustrated in Fig.\ \ref{fig_S02}. At higher metallicities the effect can also be significant,
as discussed below.
In some metal-poor, nearby galaxies, nebular continuum emission is readily observable in 
the UV-optical domain (see Fig.\ \ref{fig_guseva}), and signs of nebular emission 
are frequently observed in spectra and photometry of starburst galaxies, e.g.\ in diffuse, outer regions, or
in the near-IR domain in young star-forming galaxies such as \hii\ galaxies, blue compact
dwarfs, and similar objects \citep[cf.][]{1997ApJ...476..698I,2003Ap&SS.284..619P,2001A&A...378L..45I,2002A&A...390..481V,
2004A&A...421..519G}. 
%In these cases nebular emission (continuum and lines ) can also
%dominate the photometry of of  colors.
At high redshift a peculiar, lensed galaxy (the so-called Lynx arc at $z=3.357$) has been found, whose spectrum 
appears to be dominated by nebular continuum emission and lines \citep{2003ApJ...596..797F}.
A similar explanation has been suggested for a $z=5.5$ galaxy \citep{Raiter2010}, although alternative explanations
based on  more conventional galaxy SEDs can also be found \citep{2010A&A...513A..20V}.
In any case, the objects and conditions mentioned here represent probably just somewhat extreme
examples illustrating the potential effects of nebular emission, and their detailed importance needs
to analyzed on a case-by-case basis.
However, there is growing evidence that nebular emission plays a role for the interpretation
of observations of many/most Lyman break and \lya\ emitter galaxies at $z \ga 3$ 
\citep[cf.][and discussion below]{2011arXiv1111.6373S}.

Several evolutionary synthesis models, such as P\'EGASE, GALEV and others have
long included nebular emission (continuum emission and also lines in some cases, and 
relying on different assumptions) \cite[cf.][]{fioc99,leitherer99,Charlot01,Anders03,Zackrisson08}. 
However, other widely used codes \citep[e.g.][]{BC03} neglect this component.
To interpret emission line observations of \hii\ regions and galaxies, the stellar spectra
predicted by evolutionary synthesis codes have often been used as input for photoionisation models
\citep{1995A&AS..112...13G,1996ApJS..107..661S,2001A&A...370....1S,2001ApJ...556..121K}
%Garcia-Vargas, Stasinska+L, SS99, Kewley ....
%
Recently, nebular emission has again received considerable attention, in particular for the 
interpretation of photometric observations of distant Lyman break galaxies (LBGs) and 
\lya\ emitters (LAE) as discussed below
\citep{SdB09,SdB10,2010ApJ...724.1524O,2011arXiv1111.6373S,2011ApJ...737...47A}.
%Others: Ono Acquaviva Finkelstein

Now that all emission sources have been ``assembled", we need to briefly discuss
absorption/attenuation professes occurring along the lines-of-sight to the observer.

% % % % % % % % % % % % % % % % % % % % % % % % % % % % % % % 
\subsection{Attenuation law} 
To compare predictions of evolutionary synthesis model with observations, 
the effect of interstellar reddening (extinction and/or attenuation) obviously needs to be taken
into account. Simple prescriptions are generally used, describing e.g.\ the mean
observed attenuation law of star-forming galaxies \citep[e.g.\ the ``Calzetti law"][]{Calz01}
or various laws describing extinction in the Galaxy or the SMC. 
At very high redshift \citet{2010A&A...523A..85G} have recently argued for an attenuation law
differing somewhat from the Calzetti law, probably due to different dust composition
from supernovae dominating in the early Universe.

In the nearby Universe, stellar emission is generally found to be less attenuated than
emission lines \citet{Calz01}, which is attributed to geometrical effects.
If this also applies to distant star-forming galaxies remains, however, controversial
and may vary from case to case 
%e.g. Pettini, JAPANESE
\citep[e.g.][]{2009ApJ...701...52H,2010ApJ...718..112Y}.
Other complication arise e.g.\ in the presence of multiple stellar populations suffering from
different attenuations. More complex prescriptions and additional assumptions
are needed to describe such cases 
%see e.g. Charlot \& Fall, Granato et al. picture... 
\citep[see e.g.][]{2000ApJ...539..718C,2000ApJ...542..710G}.

%% DEGENERACIES of extinction -- age
In concrete applications of synthesis models to fitting (restframe) UV--optical photometric observations of star-forming galaxies
attenuation/extinction is often degenerate with the age of the stellar population, since both lead to
a redder SED on average. This well-known age--reddening degeneracy is illustrated e.g.\ in studies of
simple stellar populations
%SSPs
\citep[see][]{2010MNRAS.403..780C},
% distant SF galaxies: Pello et al. 1999
distant star-forming galaxies \citep{1999A&A...346..359P},
%e.g. for distant red galaxies Pozzetti, L.; Mannucci, F.
distant red galaxies \citet{2000MNRAS.317L..17P}, and in many other works.
Obviously, the use of other, more sensitive indicators of age and/or attenuation, e.g.\ 
spectral lines, the Balmer break etc.\ \citet[cf.][]{2001MNRAS.325...60C}, can reduce such degeneracies. 
A more detailed discussion of this and other issues related to the fitting of observed
SEDs and spectra is beyond the scope of this Chapter. 

% % % % % % % % % % % % % % % % % % % % % % % % % % % % % % % 
\subsection{IGM}
Finally, the collective effect of the Lyman forest on predicted spectra of distant galaxies
must also to taken into account. Indeed for redshifts above $z \ga$ 3--4 the IGM significantly
reduces the flux shortward of \lya\ in the restframe of the galaxy causing the Lyman
(continuum) break at 912 \AA\ to shift rapidly to \lya\ ($\sim$ 1216 \AA).
Generally the average IGM attenuation is described by a simple expression derived
from statistical analysis of the \lya\ forest, e.g.\ following \citet{madau95} or recent
updates of this work \citep{2008ApJ...681..831F}.

A related, important question in several contexts concerns the transmission of \lya\ photons
emitted by high redshift galaxies. Indeed, as strong \lya\ emission is one of the expected
signatures of primordial and very metal-poor stellar populations (cf.\ below), knowing its 
escape fraction out of galaxies and the subsequent transmission through the surrounding
IGM are of fundamental importance. By the same token, observations of \lya\ emission
(e.g.\ the \lya\ luminosity function of galaxies) and its evolution with redshift are used to 
constrain the IGM transmission and hence the reionization history of the Universe.
For more information on these related topics we refer the reader to a vast literature including
e.g.\ \citet{2002ApJ...576L...1H,2006ARA&A..44..415F,Schaerer07,2011ApJ...730....8H,2011MNRAS.410..830D,
2012MNRAS.tmp.2199L} and references therein.
%Lidman et al. Hayes et al. fesc(z)

%%%%%%%%%%%%%%%%%%%%%%%%%%%%%%%%%%%%%%%%%%%%%%%%%%
\begin{figure}[htb]
\sidecaption
\centering{
\includegraphics[width=0.6\textwidth]{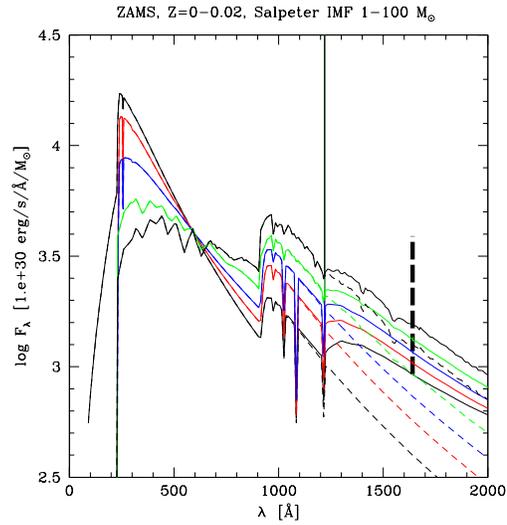}
}
\caption{Predicted SEDs including \lya\ and \Heiiuv\ emission lines
for zero age main sequence (ZAMS) models at different metallicities.
The metallicities $Z=$ 0. (Pop III), $10^{-7}$, $10^{-5}$, 0.0004, and 
0.02 (solar) are from top to bottom in the EUV ($\lambda <$ 912 \AA),
and reversed at longer wavelengths. The dashed lines are the pure
stellar emission, the solid lines show the total (stellar $+$ nebular) 
emission. A Salpeter IMF from 1--100 \msun\ is assumed here for all metallicities.
From \citet{Scha03}.}
\label{fig_sed_s02}
\end{figure}

\begin{figure}[htb]
\sidecaption
\centering{
\includegraphics[width=0.6\textwidth]{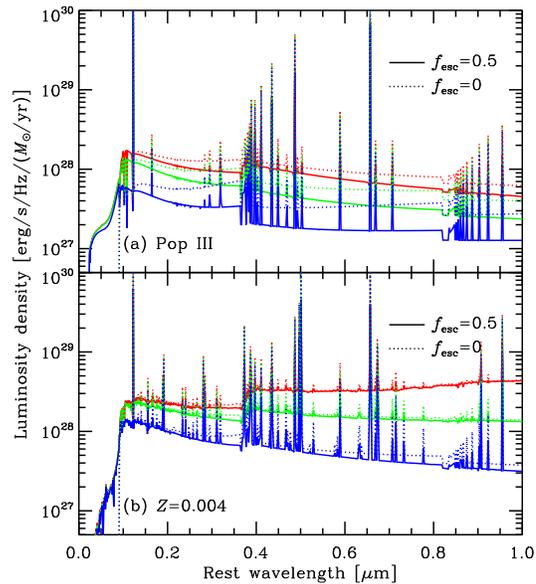}
}
\caption{Predicted model spectra of  a PopIII galaxy (top panel) and a moderate sub-solar metallicity ($Z = 1/50 \zsun$,
bottom panel), at ages of 10, 100, and 500 Myr (from bottom to top in each panel), evolving with a constant SFR. 
Solid and dotted lines show cases computed for 50\% and no escape of Lyman continuum photons.
Figure from \citet{2011MNRAS.415.2920I}.}
\label{fig_inoue_sed}
\end{figure}

%%%%%%%%%%%%%%%%%%%%%%%%%%%%%%%%%%%%%%%%%%%%%%%%%%
\section{From present-day metallicities back to the first galaxies}
\label{s_predict}

The {\em Starburst99} evolutionary synthesis models \citep{leitherer99} and the models
of \citet{Scha02,Scha03} allow one to examine in detail the dependence of the expected observational
properties of star-forming galaxies with metallicities from current (solar) metallicities to metal-poor
and metal-free cases. Furthermore \citet{Scha02,Scha03} and \citet{Raiter2010} also present calculations for a 
variety of different IMFs, such that IMF or combined IMF/metallicity changes can be examined.
We here summarize the behavior of some of the main observables based on these models. 

% % % % % % % % % % % % % % % % % % % % % % % % % % 
\subsection{UV--optical: stellar and nebular continuum emission}
\label{s_uvslope}
Figure \ref{fig_sed_s02} shows variations of the UV spectrum, including the ionizing ($<912$ \AA) and 
non-ionizing part ($\lambda >$ 912 \AA) of a young stellar population with metallicity. Three main 
features are immediately clear:
First, due to the increase of the average stellar temperature, the stellar spectrum becomes harder/bluer with decreasing metallicity $Z$.
Second, the contribution of nebular continuum emission increases concomitantly. At young ages, the continuum longward
of \lya\ ($>1216$ \AA) will be dominated by nebular emission. Finally, H and He emission lines become also stronger
with decreasing $Z$ (cf.\ below).
The models in Fig.\ \ref{fig_sed_s02}  assume so-called case B recombination, i.e.\ in particular that all ionizing photons
are absorbed within the surrounding \hii\ region. This case obviously maximizes the nebular emission, whereas 
in a more general case where a fraction \fesc\ of Lyman continuum photons escape, nebular emission will be 
decreased; see e.g.\ \citet{2011MNRAS.415.2920I} and Fig.\ \ref{fig_inoue_sed}.

\begin{figure}[htb]
\sidecaption
\centering{
\includegraphics[width=0.6\textwidth]{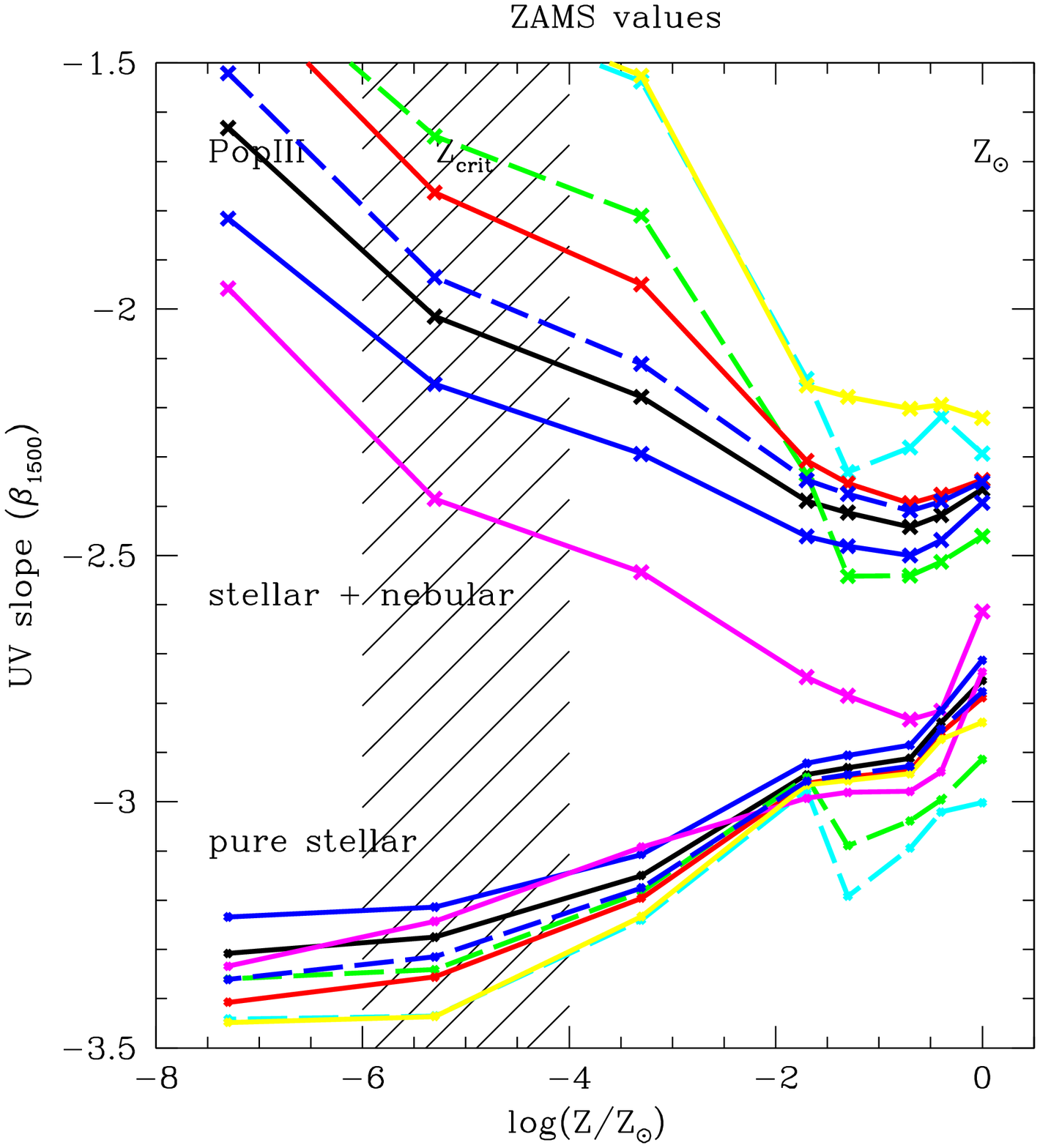}
}
\caption{Predicted UV slope $\beta_{1500}$ for different IMFs (color codes) plotted as a function metallicity. The values are shown for very young (ZAMS) populations, which correspond to the bluest possible slopes (i.e.\ minimal $\beta$ values).
The upper set of lines shows the UV slopes of the total spectrum (stellar + nebular continuum),
the lower lines using the pure stellar spectrum. 
Black lines show the predictions for a Salpeter IMF slope. All color codes are given in
\citet{2010A&A...523A..64R}, from where the figure is taken.}
\label{fig_beta_zams}
\end{figure}

\begin{figure}[htb]
\sidecaption
\centering{
\includegraphics[width=0.6\textwidth]{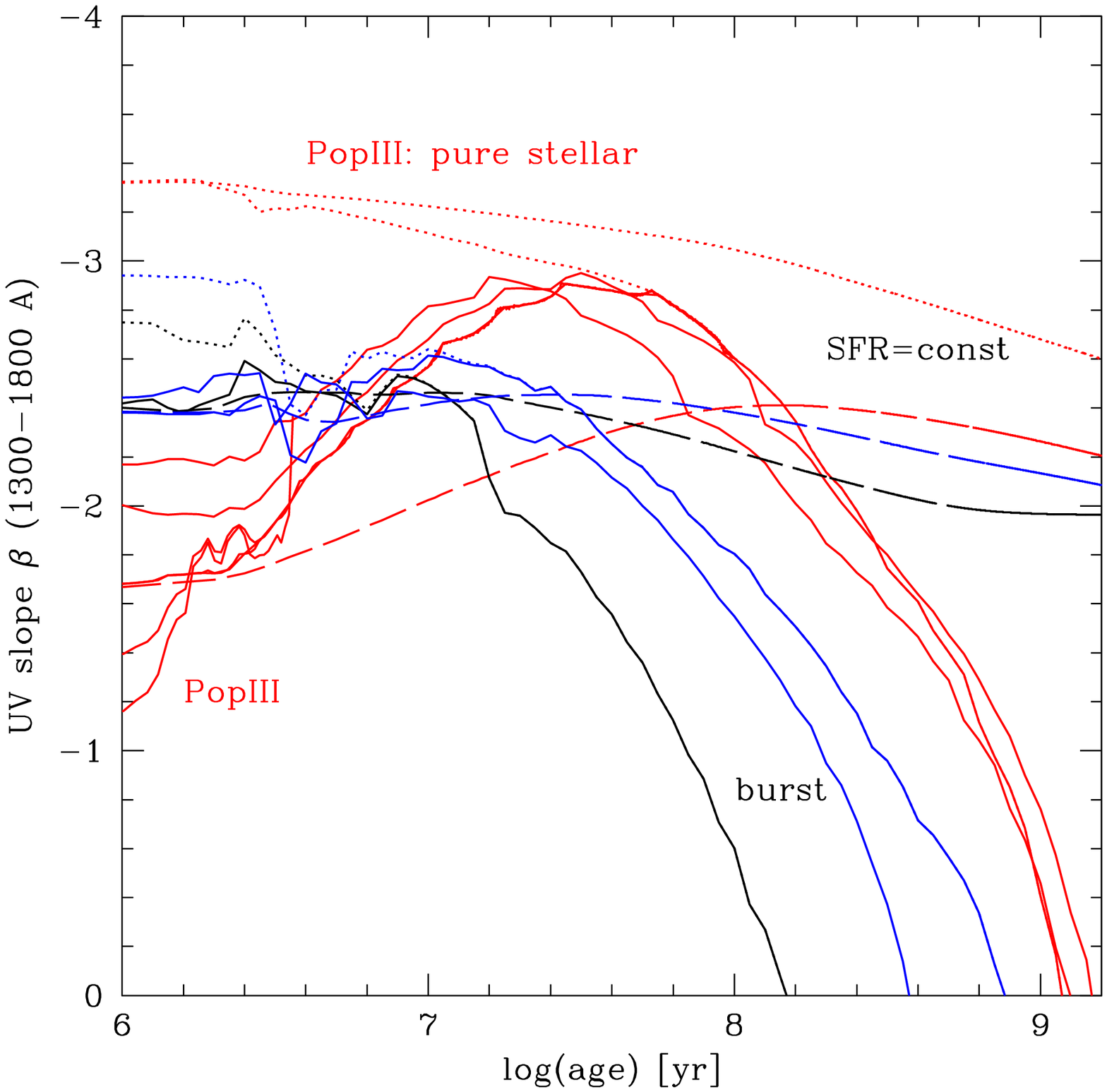}
}
\caption{Temporal evolution of the UV slope $\beta_{1500}$ 
%measured between 1300 and 1800 \AA\ 
derived from synthesis models of different metallicities and for instantaneous bursts (solid lines) and constant SF (long dashed lines). Black lines show solar metallicity models, red lines show metallicities between $Z = 10^{-5}$ and zero (PopIII) and blue lines show intermediate cases of $Z = 0.004$ and 0.0004. The dotted lines show $\beta$ if nebular continuous emission is neglected, that is, assuming pure stellar emission. Note especially the strong degeneracies of $\beta$ in age and metallicity for bursts, the insensitivity of $\beta$ on Z 
for constant SF, and the rather red slope for young very metal-poor bursts. From \citet{Schaerer2005}.}
\label{fig_beta}
\end{figure}

The color of the UV continuum and its variations can be quantified e.g.\ by the exponent $\beta$ of 
a power law adjusted to a specific spectral region. This so-called ``$\beta$-slope"
determined from models and measured by spectroscopy or photometry is often used in the literature,
in particular to estimate the amount of dust attenuation in high-z star-forming galaxies
\citep[see e.g.][and also the Chapter by Dunlop in this book]{1999ApJ...521...64M,Bouwens09_beta}.

Figure \ref{fig_beta_zams} shows the dependence of the predicted $\beta$-slope
(here measured between 1300 and 1800 \AA) of a young (zero age) population on metallicity and on 
the IMF (coded by different colors). Fig.\ \ref{fig_beta} shows the same quantity as function
of age for instantaneous bursts and constant SFR.
From Fig.\ \ref{fig_beta_zams} we see how the {\em stellar spectrum} steepens (becomes bluer) 
with decreasing $Z$. However, once nebular emission is taken into account (here with \fesc=0) the UV
spectrum becomes significantly flatter and the trend with metallicity even inverses!
In short, the UV slope cannot be a good metallicity indicator.
Figure \ref{fig_beta} shows the rapid evolution of $\beta$ on timescales of $\sim$ 50--100 Myr.
For constant SFR, $\beta$ reaches the typical asymptotic value of $\beta \sim -2$ to $-2.4$,
the precise value depending on the wavelength base used to define $\beta$.  
Shifts of $\sim$ 0.1--0.3 in $\beta$ can be typical, as illustrated e.g.\ by \citet{2010A&A...523A..64R}.
For PopIII we see that the nebular contribution vanishes after $\ga 20$ Myr in a burst;
between ages of $\sim$ 10--100 Myr an integrated population can reach a very blue spectrum
with $\beta \sim -2.5$--3. Otherwise, and for more extended periods of star formation it is difficult,
if not impossible, to obtain a UV spectrum steeper than $\beta \sim -2.5$, except if a significant
fraction of Lyman continuum photons are ``leaking'' (i.e.\ $\fesc >0$). 
Overall, the two figures show in particular the following: 
First, in principle metallicity cannot be inferred from the 
observed UV slope \citep[cf.][]{Schaerer2005}.
Second, very steep/blue UV spectra ($\beta \ll -2.5$) are not predicted for very metal-poor, primordial, populations,
except if nebular emission is ``suppressed".

%Unusually blue UV slopes have e.g.\ recently been claimed by \citet{Bouwens10_betaz7}.
Some authors have indicated the possibility of unusually blue UV slopes in $z \sim 7$ galaxies \citep{Bouwens10_betaz7}.
However, the significance of these findings is low \citep{SdB10,2010ApJ...719.1250F},
and independent and more recent measurements \citep{2011arXiv1109.1757C,2011arXiv1109.0994B,2012MNRAS.420..901D}
do not show indications for exceptional populations or conditions (PopIII and strong leakage of Lyman continuum 
photons).  Other observational aspects related to the UV slope of distant star-forming galaxies
are discussed in the Chapter of Dunlop.

% % % % % % % % % % % % % % % % % % % % % % % % % % 
%\newpage
\subsection{Ionizing photon production}
\begin{figure}[htb]
\sidecaption
\centering{
\includegraphics[width=0.6\textwidth]{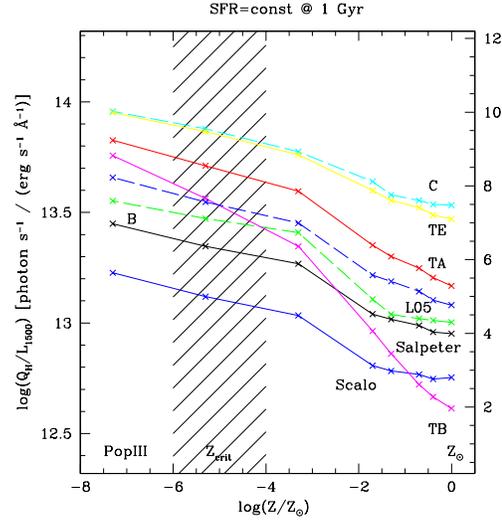}
}
\caption{Predicted Lyman continuum flux as function of metallicity.
The relative output of hydrogen ionizing photons to UV continuum light, measured
at 1500 \AA\ restframe, $Q_H/L_{1500}$, is shown as a function of metallicity 
for constant star formation over 1 Gyr.
%(left panel) and very young bursts (right panel).
$Q_H/L_{1500}$ is given in $L_\lambda$ units on the left side of each
panel, and in $L_\nu$ units on the right.
Results for different IMFs are shown using the color codes and labels
summarised in Table 1 of  \citet{2010A&A...523A..64R}.
The shaded area indicates the critical metallicity range where
the IMF is expected to change from a ``normal'' Salpeter-like regime to   
a more massive IMF (see text).
From  \citet{2010A&A...523A..64R}.}
\label{fig_qh}
\end{figure}

To quantity the contribution of galaxies to cosmic reionization, and to predict their observable spectra 
it is important to know the amount of energy or photons emitted in the Lyman continuum,
i.e.\ at $\lambda < 912$ \AA\ or at energies $E>13.6$ eV.
This quantity is straightforwardly predicted by evolutionary synthesis models, and is generally
expressed as a photon flux (e.g.\ in units of photons per second) normalized for example
per unit stellar mass, per unit SFR, or so. Another, observationally relevant way to express the Lyman
continuum flux is by normalizing it to the UV output of the same stellar population, since the latter
is a direct observable. This prediction is shown, for a constant star formation rate over relatively
long timescales, as a function 
of metallicity and for different IMFs in Fig.\ \ref{fig_qh}, taken from \citet{2010A&A...523A..64R}.

As expected $Q_H/L_{1500}$ increases with decreasing metallicity,
since the ionizing flux depends very strongly on the effective
stellar temperature and hence increases more rapidly than the UV
luminosity. The IMF dependence also behaves as expected, with the 
IMFs favouring the most massive stars showing also the highest
the $Q_H/L_{1500}$ ratios, since $Q_H$ increases more rapidly with
stellar mass than the UV luminosity.
For constant SFR and for a fixed IMF, the increase of the relative ionizing
power from solar metallicity to Pop~III is typically a factor 2 to 3.
When considering an IMF change from Salpeter to a massive IMF
(i.e.\ all cases except Salpeter and Scalo) the increase of  $Q_H/L_{1500}$ 
is larger, approximately 0.6 to 1 dex between solar and zero metallicity.
For a very young (zero age) population the Lyman continuum production increases
typically by \lsim 40 \%.

Obviously, the ionizing photon output also depends on age and the assumed
star formation history, not illustrated here \citep[see e.g.][where other normalisations are also used]{Scha02,Scha03,2010A&A...523A..64R}.
For example, for zero age stellar populations the ionizing photon output
per UV flux, $Q_H/L_{1500}$, is higher than shown here, typically by up to a factor 2--4
depending on the IMF  \citep[see Fig.\ 1 of][]{2010A&A...523A..64R}.
For very ``massive" IMFs (i.e.\ IMFs favoring very massive stars) the predictions
shown in Fig.\  \ref{fig_qh} for constant star formation over long timescales
(up to 1 Gyr) are of course very similar to those for much shorter, probably more realistic timescales
for the first galaxies, since in any case the lifetimes of the bulk of the stars is much
shorter than 1 Gyr.
For general ages and star formation histories these quantities can be derived
from the available data files of the models.

% % % % % % % % % % % % % % % % % % % % % % % % % % 
\subsection{\lya\ emission}
\label{s_lya}

Given predictions for the Lyman continuum flux from a population, the flux in the different H and He recombination lines
and nebular continuum emission from the surrounding \hii\ region can easily be computed, as described
above (Sect.\ \ref{s_nebular}), using simple case A or B recombination theory or computing detailed photoionization models.
The equivalent widths of these lines -- a very useful measure of their strength -- can also be computed, 
given the stellar and nebular continuum.

\begin{figure}[htb]
\sidecaption
\centering{
\includegraphics[width=0.6\textwidth]{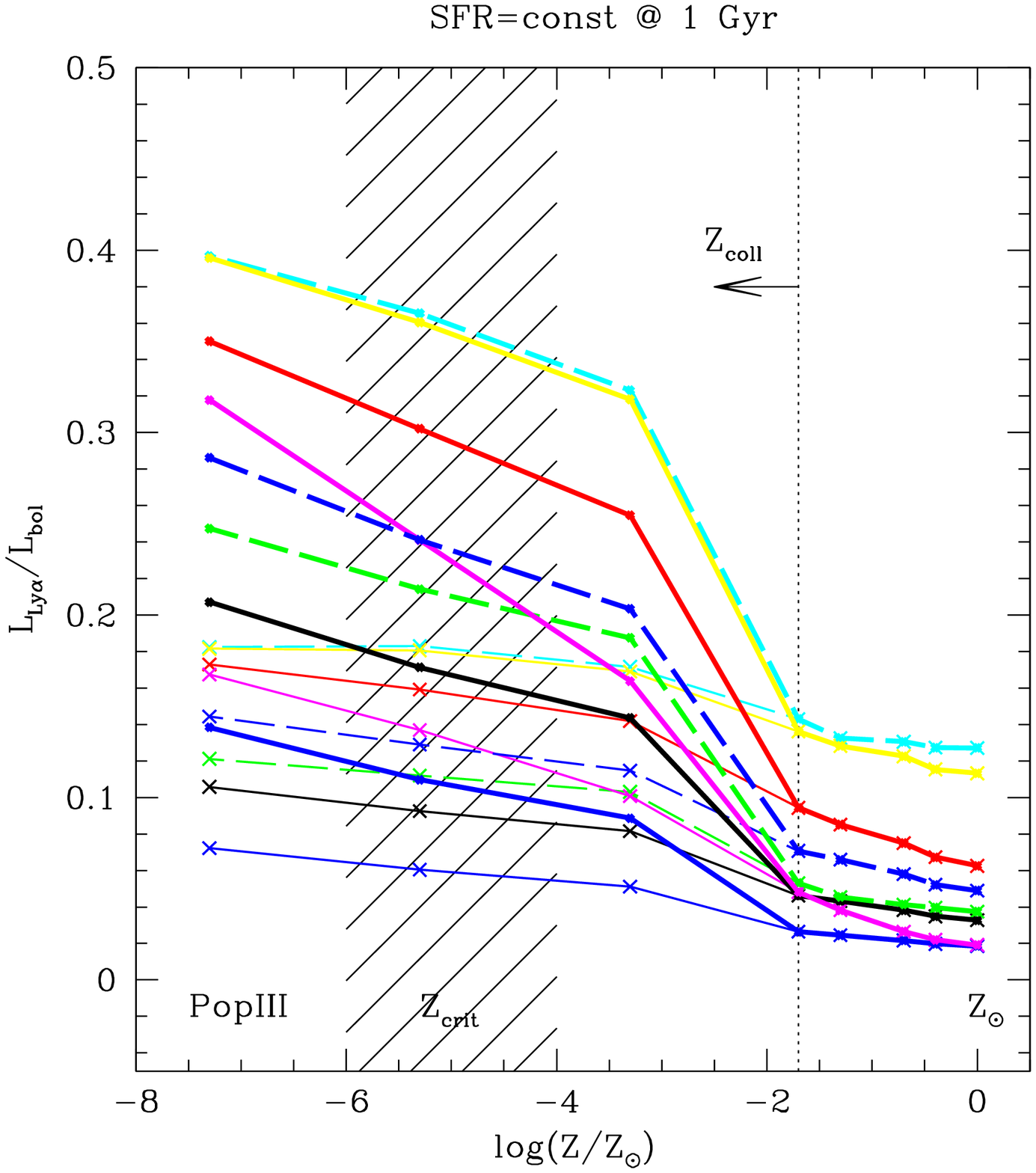}
}
\caption{Fraction of the \lya\ luminosity to the total bolometric luminosity,
$L(\lya)/L_{\rm bol}$ for SFR=const as a function of metallicity and IMF. 
Results for different IMFs are shown using the same colour codes as in
Same color codes as in Figs.\ \protect\ref{fig_beta_zams}, \protect\ref{fig_qh}.
Thin lines show the results using standard case B recombination; thick lines the 
recent results accounting for departures from case B at very low
metallicity.  Note the strong increase of the predicted $L(\lya)/L_{\rm bol}$ values
from solar to very low metallicity.
From  \citet{2010A&A...523A..64R}.}
\label{fig_lyafraction}
\end{figure}

Since intrinsically the strongest line, and conveniently located in the rest frame UV domain, the \lya\ line
is well known to be central to many studies of distant/primeval galaxies, in particular since the early work of \citet{pp67}
and the discovery of large populations of high redshift galaxies \citep[e.g.][and many other papers on such observations]{Hu98,Ouc08}.
Before the era of synthesis models and based on relatively crude assumptions, \citet{pp67} estimated that up to $\sim$ 10\% 
of the bolometric luminosity of primeval galaxies could be emitted in the \lya\ line.
For constant star formation the most recent update, using our current knowledge of stellar evolution, atmospheres and the Salpeter
IMF, place this number at $\sim$ 3 \% for solar metallicity, as shown in Fig.\ \ref{fig_lyafraction}.
%A significantly higher fraction of the bolometric luminosity is 
For the same IMF this fraction is expected to be significantly higher at low metallicity, especially below
$Z$ \la $1/50$ \zsun, where departures from case B (due to collisional effects because of the high electron temperature
in the nebula) significantly increase the ionization and hence also recombination rates, as recently
showed by \citet{2010A&A...523A..64R}. As illustrated in Fig.\ \ref{fig_lyafraction}, the \lya\ line can therefore
carry up to $\sim$ 20--40\% (depending on the IMF) of the bolometric luminosity in primordial gas.
As already discussed above for the ionizing photon flux, the precise fraction of the \lya\ luminosity
obviously also depends on the age and star formation history. In any case, for the most extreme cases
shown here (with IMFs dominated by massive stars), the predictions in Fig.\ \ref{fig_lyafraction} 
do not vary much with age, since the bulk of stars in such populations have anyway short lifetimes.

\begin{figure}[htb]
\sidecaption
\centering{
\includegraphics[width=0.6\textwidth]{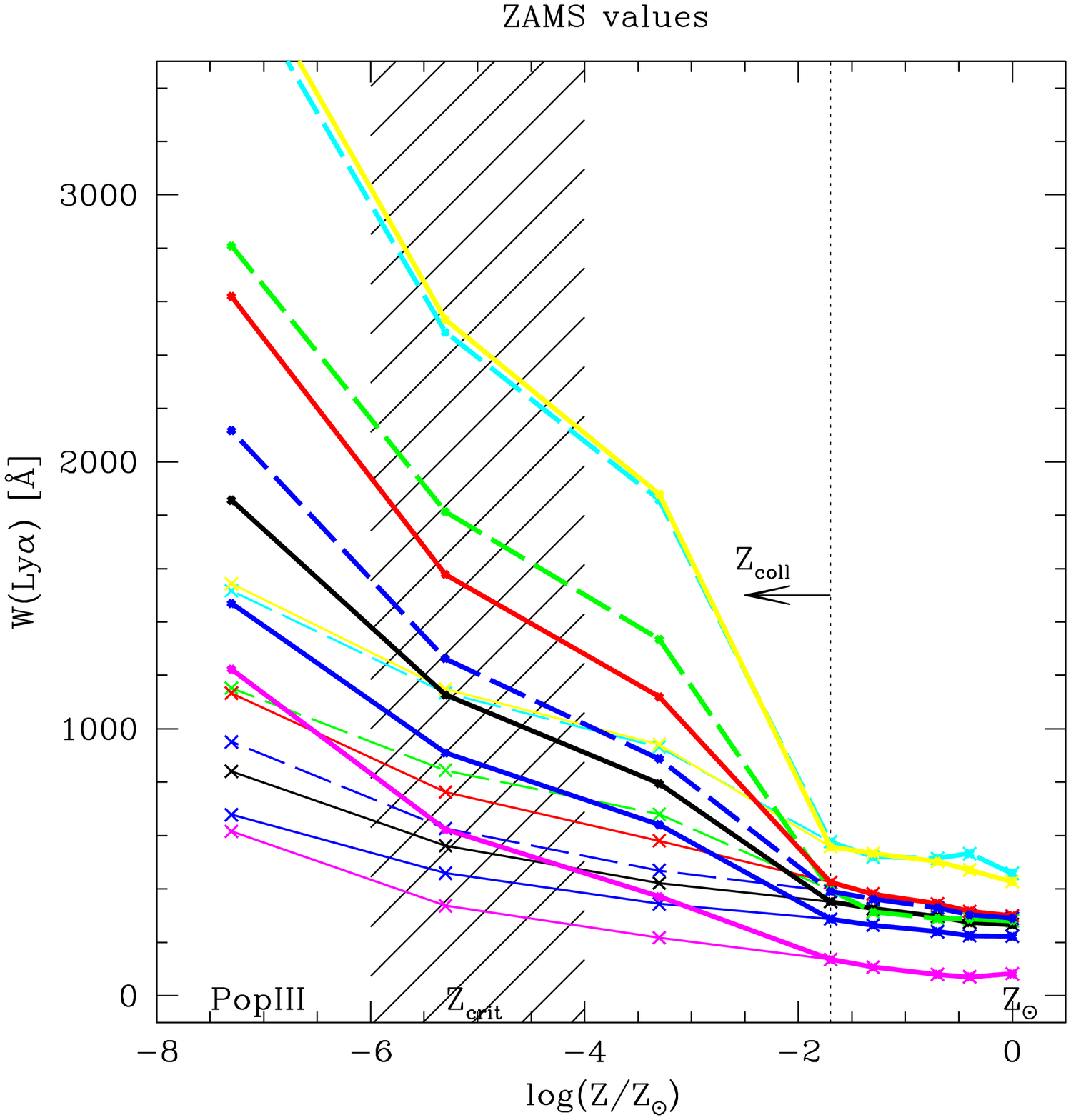}
}
\caption{Predicted \lya\ equivalent width as a function of metallicity 
for very young ($\le$ 1--2 Myr).
Thin lines show  standard case B predictions, thick
lines the predicted $W(\lya)$ accounting to first order for departure from case B,
%following Eq.\ \protect\ref{eq_lya_final} (assuming low density, i.e.\ 
%$\tilde{f}_{\rm coll}=2/3$, and neglecting the increase of the two-photon continuum), 
leading to an increase by up to a factor $\sim$ 1.5--2.5 at low metallicities 
($Z \protect$\lsim$\zcoll$). %
%Results for different IMFs are shown using the same colour codes as in
%Fig.\ \ref{f_nlyc} (cf.\ Table \ref{t_imf}).
Same color codes as in Figs.\ \protect\ref{fig_beta_zams}, \protect\ref{fig_qh},
\protect\ref{fig_lyafraction}.
From  \citet{2010A&A...523A..64R}.}
\label{fig_ewlya}
\end{figure}

The maximum strength of \lya, measured by its equivalent width \wlya, is recognized as an interesting
diagnostic of young, metal-poor/metal-free stellar populations in the first galaxies. Indeed, very high 
equivalent widths -- well beyond the maximum \wlya\ of $\sim$ 200--250 \AA\ predicted for solar metallicities --
are expected for such populations. As such, or together with other unique spectral features of \heii\ discussed below, 
this line is of great interest to search for ``unusual" stellar populations \citep[see e.g.][]{Malhotra02}.
While various initial predictions of the \lya\ equivalent widths have yielded somewhat different results
\citep[see e.g.][]{Tumlinson01,Bromm01}, it is now clear that non-LTE atmosphere models, the treatment of nebular
emission, and departures from case B, are essential ingredients to properly predict the strength and equivalent
widths of \lya\ emission from very metal-poor populations 
\citep[see][]{Scha02,Scha03,2010A&A...523A..64R}. 

With these ingredients, the maximum \wlya\ predicted
as a function of metallicity and for different IMFs reaches $\sim$ 2000 \AA\ in the rest-frame (for a Salpeter IMF),
or higher for IMFs favoring more massive stars, as illustrated in Fig.\ \ref{fig_ewlya}.
In principle, observations of such large equivalent widths of \lya, if attributable to photoionization from 
stars (as opposed to non-thermal sources), should be strong sign-posts of extreme conditions (metallicity
and/or IMFs) expected in the first galaxies.
Possible complications for the application of such a diagnostic include the effect of the IGM and dust/radiation transfer
inside galaxies, which can significantly reduce the \lya\ flux (and more so than the nearby UV continuum), and hence
reduce \wlya. Furthermore the  \lya\ line emission varies rapidly with time (on timescales of $\sim 10^7$ Myr).
\citet{neufeld91} has suggested that radiation transfer in a clumpy ISM could increase \wlya, which, if applicable,
could complicate the interpretation of high \wlya\ objects. In any case it is clear that galaxies with
unusually high \lya\ equivalent widths (\ga\ 200--250 \AA) are interesting candidates worth examine further 
in searches for very metal-poor and PopIII galaxies.

% % % % % % % % % % % % % % % % % % % % % % % % % % 
\subsection{Hardness of the ionizing spectrum and Helium line emission from the first galaxies}

As already mentioned above, the first stellar populations are expected to contain unusually hot massive stars
whose ionizing spectra will be harder than that of ``normal" present-day massive stars. As a consequence
these stars will emit in particular more ionizing photons above 54 eV ($\lambda <$ 228 \AA), the energy required 
to fully ionize Helium in the ISM surrounding these stars. This will give rise to recombination lines of \heii\
observable in the UV and optical domain (e.g.\ \Heiiuv, \Heiiopt), which should therefore be a fairly
unique signature of these energetic, metal-poor stars.

\begin{figure}[htb]
\sidecaption
\centering{
\includegraphics[width=0.6\textwidth]{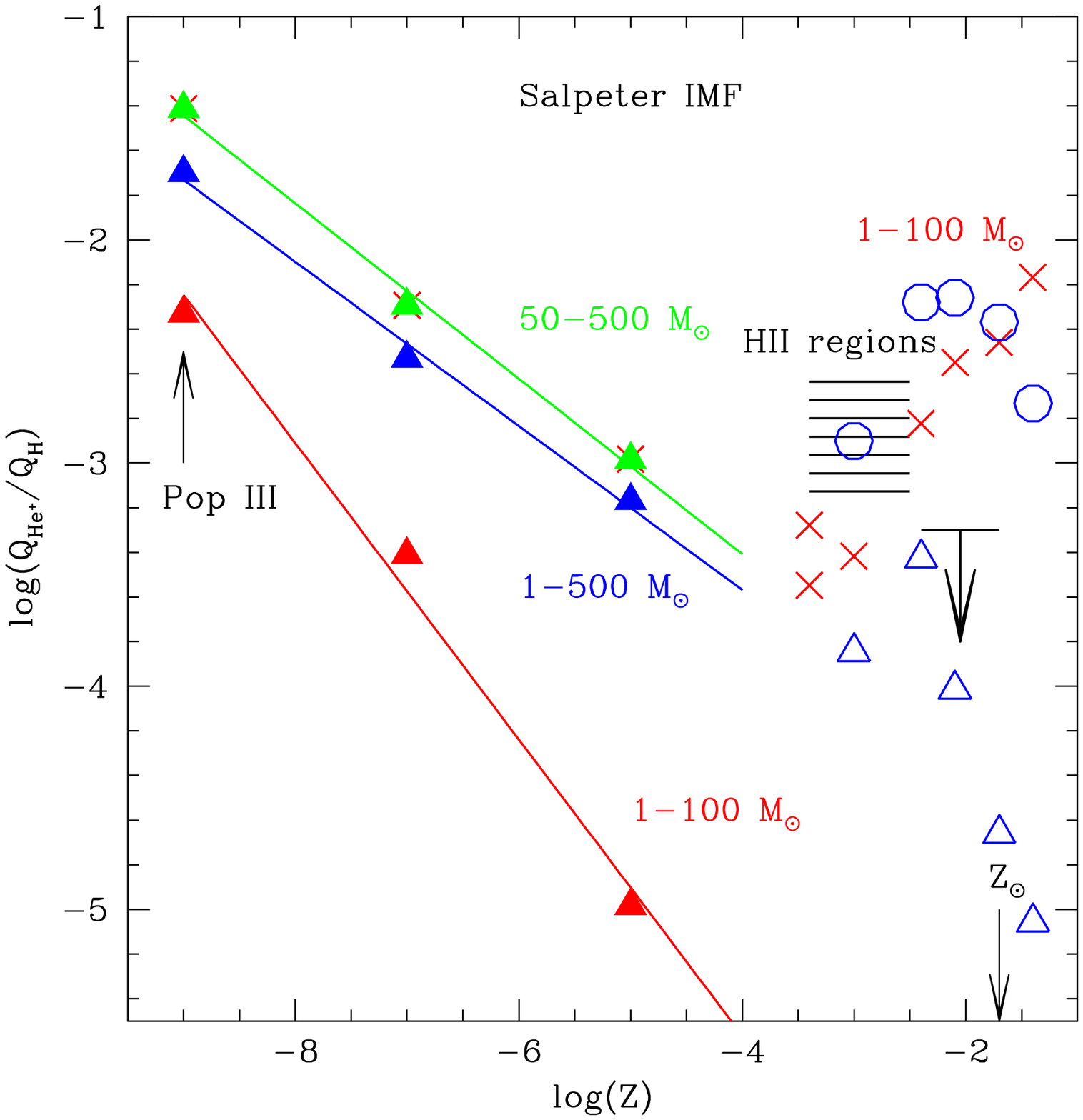}
}
\caption{Hardness \Qrathep\ of the He$^+$ ionising flux for constant star 
formation as a function of metallicity (in mass fraction) 
for three different power-law IMFs.
%all models given in Table \ref{tab_csf}.
At metallicities above $Z \ge 4. \, 10^{-4}$ the predictions from
our models (crosses), as well as those of Leitherer \etal\ (1999, 
open circles), and Smith \etal\ (2002, open triangles) are plotted.
The shaded area and the upper limit (at higher $Z$) indicates 
the range of the empirical hardness estimated from 
\hii\ region observations \citep[see discussion in ][]{Scha03}.
From \citet{Scha03}.}
\label{fig_hardness}
\end{figure}

\begin{figure}[htb]
\sidecaption
\centering{
\includegraphics[width=0.6\textwidth]{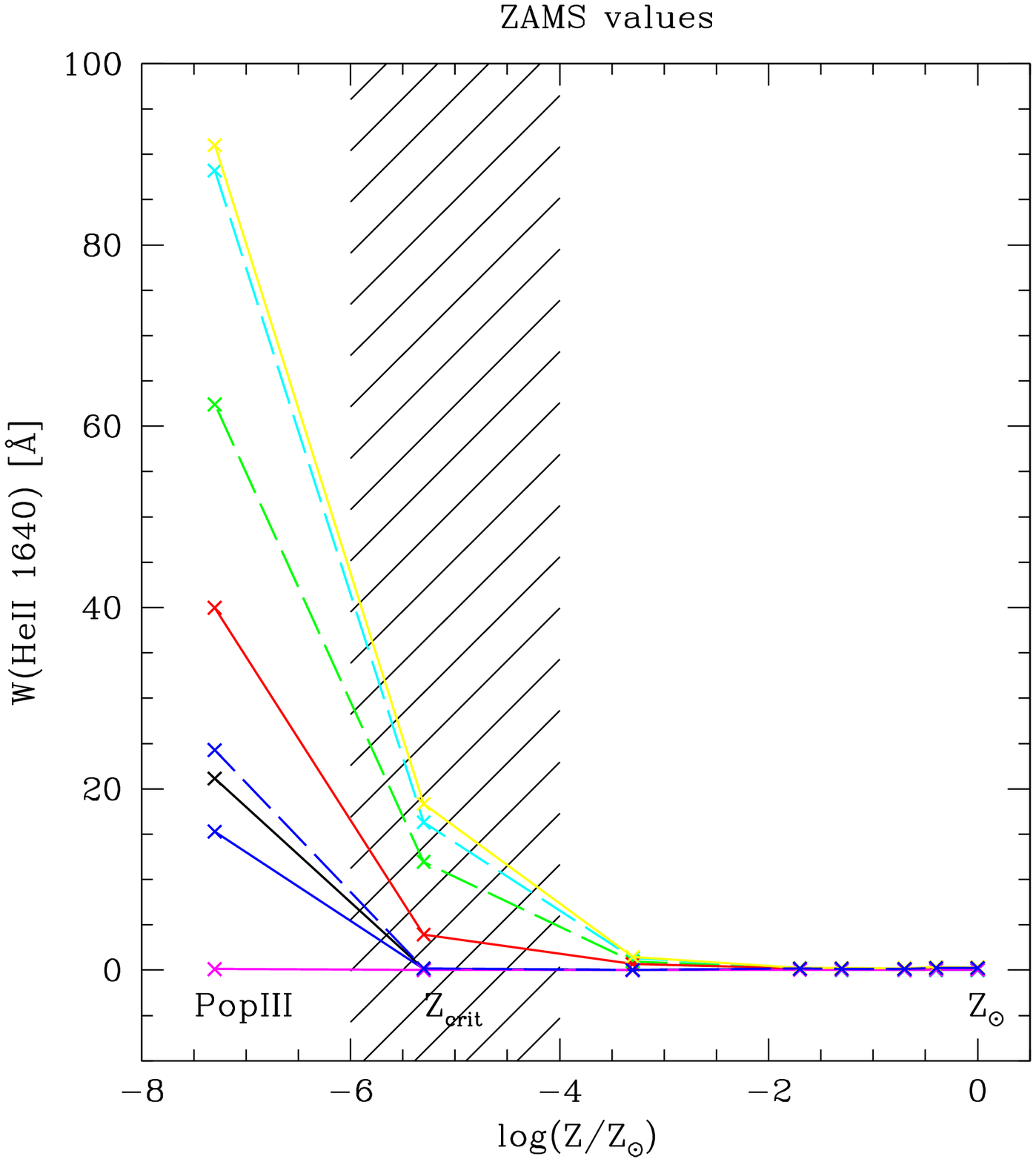}}
\caption{Predicted \Heiiuv\ equivalent width as a function of metallicity 
for very young ($\le$ 1--2 Myr) bursts.
Same color codes as in Figs.\ \protect\ref{fig_beta_zams}, \protect\ref{fig_qh},
\protect\ref{fig_lyafraction}.
Note that photoionization models predict generally fainter \Heiiuv\
emission, hence lower equivalent widths, except for high ISM densities.
From \citet{2010A&A...523A..64R}.}
\label{fig_heiiuv}
\end{figure}

The use of \heii\ lines to identify PopIII stars/galaxies was discussed by 
\citet{Tumlinson00,Tumlinson01,Bromm01,2001ApJ...553...73O,Scha02,Scha03}, who 
present predictions for the strength of these lines. \citet{Scha03} and more recently \citet{2010A&A...523A..64R}
have discussed the transition from zero metallicity to present-day conditions. The predicted hardness
of the ionizing flux of starbursts at different metallicities and the behavior of the \Heiiuv\ line as an example,
are shown in Figs.\ \ref{fig_hardness} and \ref{fig_heiiuv}.
The main result is indeed a strong increase of the hardness of the ionizing flux, typically by 2 orders
of magnitudes from metallicities $Z \sim$ 1/50 \zsun\ to PopIII (zero metallicity).
The (maximum) equivalent width of \Heiiuv\ (again taking the ``dilution'' by both the stellar and nebular continuum into account)
is generally very low. Values above $\ga$ 5 \AA\ (rest frame) are only expected for young, very metal-poor stellar 
populations. One should, however, remember that these predictions are based on simple assumptions
(case B, ionization bounded regions), which may not always apply.

\citet{2009MNRAS.399...37J} have computed the emission in H and \heii\ recombination lines from
hydrodynamic simulations taking into account the (time-dependent) leakage of Lyman continuum and He$^+$ ionizing photons.
They find a smaller leakage for the higher energy photons, as expected, since for stellar sources these less abundant
photons are absorbed closer to the source than H ionizing photons. They therefore conclude that the \heii\ equivalent 
width should indeed be a fairly robust indicator for PopIII.

As for \lya, the prediction of the intrinsic emission in Helium lines may also be more complicated than expected from
simple ``photon-counting''  assumed for case B (and implemented in most synthesis models). The basic reason for this
is that both H and He in the ISM compete for ionizing photons, which leads to a lower He$^+$ ionization rate in regions 
of low ionization parameter. This effect, already discussed by  \citet{Stasinska86} for planetary nebulae, implies that the
\heii\ line flux may be lower than predicted by evolutionary synthesis models, as shown by \citet{2010A&A...523A..64R}.
Detailed photoionization models are necessary to properly account for this effect.

\begin{figure}[htb]
\sidecaption
\centering{
\includegraphics[width=0.8\textwidth]{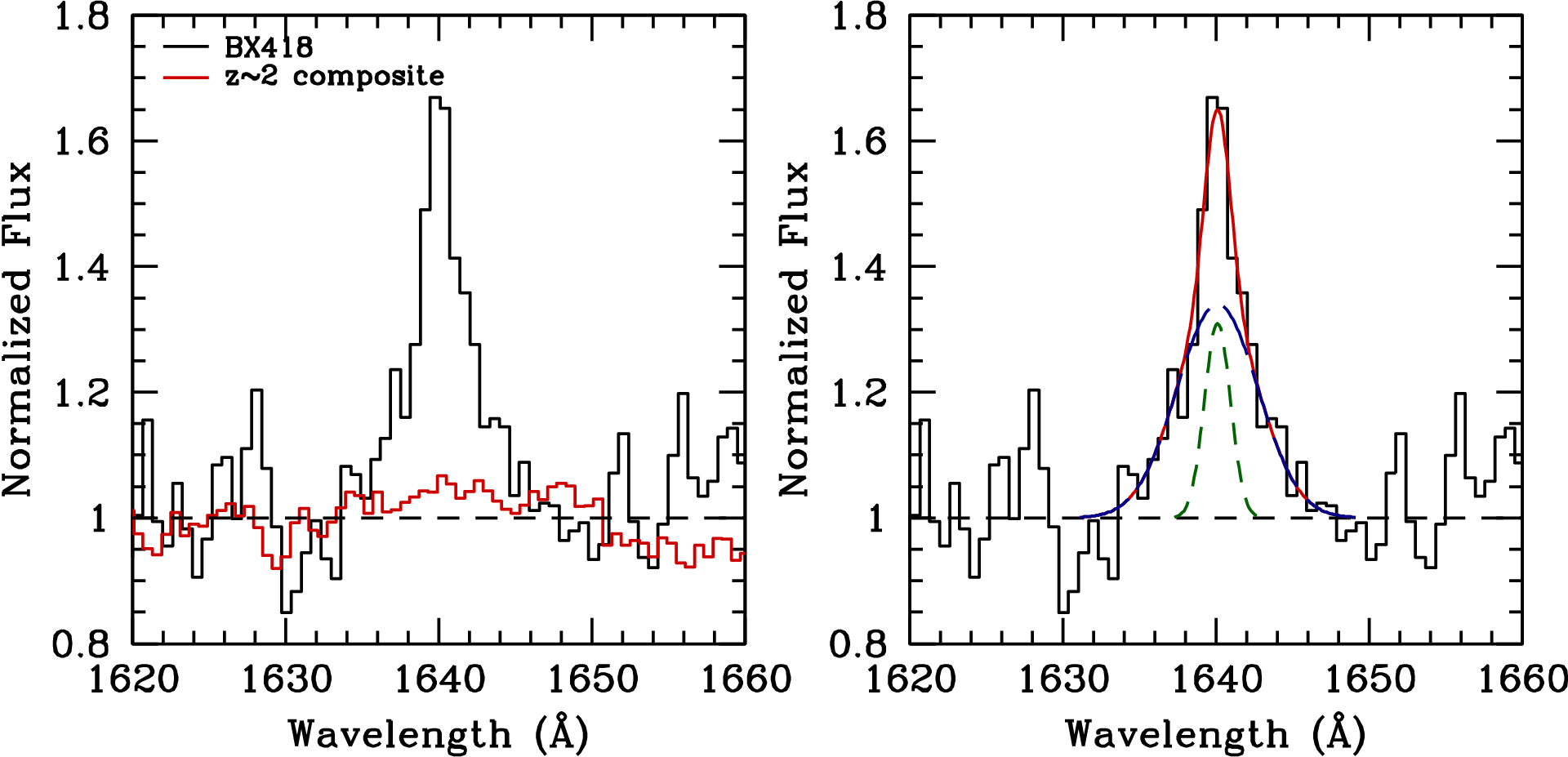}}
\caption{{\em Left:} \Heiiuv\ emission in the peculiar $z\sim 2$ galaxy BX418 (black), compared to the broad and far weaker emission in the composite 
spectrum of 966 $z \sim 2$ galaxies (red). {\em Right:} The line  is well fit by a superposition of two Gaussian components (solid red line). The broad component (long-dashed dark blue line) has FWHM $\sim$ 1000 \kms, while the narrow component (short-dashed green line) is unresolved. The broad emission is attributed to W-R stellar winds, and the narrow component to nebular \heii\ emission. 
The line has a total equivalent width of 2.7 \AA, considerably smaller than expectations for very metal-poor stellar populations.
Figure from \citet{2010ApJ...719.1168E}.}
\label{fig_wrline}
\end{figure}

Finally -- as always -- it is also useful and important to examine what is known empirically about \heii\ emission.
As well known, some stars (mostly the so-called Wolf-Rayet stars, evolved massive stars) show emission in \heii\ 
lines, and these lines are visible in the integrated spectra of some galaxies 
\citep[sometimes eluded to as Wolf-Rayet galaxies, cf.][]{1991ApJ...377..115C,1999A&AS..136...35S,2008A&A...485..657B}.
The \Heiiuv\ line is also seen in some spectra (individual or stacked) of high redshift galaxies \citep[e.g.][]{Shap03,2010ApJ...719.1168E},
as shown in Fig.\ \ref{fig_wrline}. The observed equivalent widths remain, however, small compared to the values
expected at very low metallicities.
Due to the strong winds in the atmospheres of Wolf-Rayet stars these lines are broad, extending typically up to several thousand
km/s \citep{2008flhs.book.....C}. In spectra of sufficient signal-to-noise and spectral resolution it should therefore
be possible to separate any stellar emission from nebular \heii\ emission (cf.\ Fig.\ \ref{fig_wrline}).
Nebular \heii\ emission is observed in some low-metallicity \hii\ regions and starburst galaxies in the nearby/low-$z$ Universe
\citep[see e.g.][for a compilation and references therein]{1999A&AS..136...35S}. The hardness of the ionizing flux
inferred from the relative \Heiiopt/\hb\ intensities is shown by the shaded region in Fig.\ \ref{fig_hardness}; it is lower
by $\sim$ 1--1.5 orders of magnitude than what is expected for PopIII dominated objects!
However, the origin of nebular \heii\ emission in nearby objects remains difficult to understand in many
cases and several sources/mechanisms may contribute to it
\citep[see e.g.][]{1996ApJ...467L..17S,2001A&A...378L..45I,2004ApJ...606..213T,2011A&A...526A.128K,2012arXiv1201.1290S}.

Despite these open questions, the prediction of very hard spectra from primordial stars and stellar populations remains
quite solid, with the main uncertainty probably being the IMF of these stars. Signatures from \heii\ emission
should therefore be a crucial tool for observational searches of the first stellar generations.

%%%%%%%%%%%%%%%%%%%%%%%%%%%%%%%%%%%%%%%%%%%%%%%%%%
%\newpage
\section{The main observables and how to distinguish Population III?}
\label{s_obs}

From the predictions from evolutionary synthesis models discussed in the previous Section it is already apparent how PopIII
or very metal-poor populations can be distinguished from those of ``normal"  metallicities, mostly using H and He recombination
lines. Other diagnostics can be derived from integrated colors, at least to some extent. Obviously, direct
measurements of metallicity are a third way to tackle this question observationally. We here briefly summarize/describe
these various diagnostics.

% % % % % % % % % % % % % % % % % % % % % % % % % % 
\subsection{Hydrogen and Helium lines}
The predicted behavior of recombination lines of H and He$^+$ with metallicity has already been discussed above.
In short, unusually strong \lya\ emission (say \ewlya $\gg$ 250 \AA) and/or strong \heii\ emission are among the 
best indicators expected for very metal-poor stellar populations, due to their unusually high temperatures
plus possibly a more ``massive" IMF (i.e.\ favoring more massive stars).

% % % % % % % % % % % % % % % % % % % % % % % % % % 
\subsection{Metal lines}

\begin{figure}[htb]
\sidecaption
\centering{
\hspace*{-0.05\textwidth}
\includegraphics[width=0.5\textwidth]{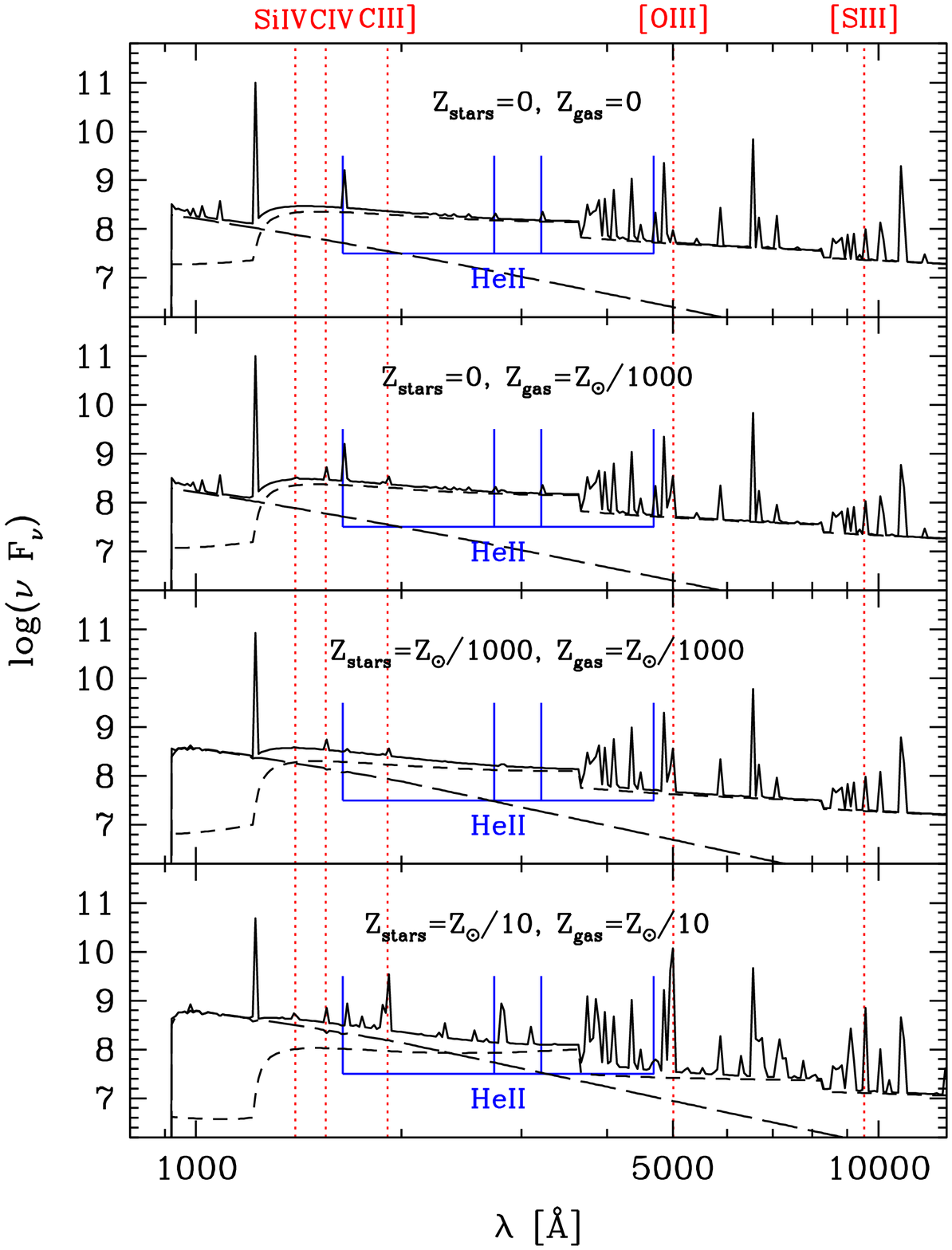}
\hspace*{0.05\textwidth}
\includegraphics[width=0.5\textwidth]{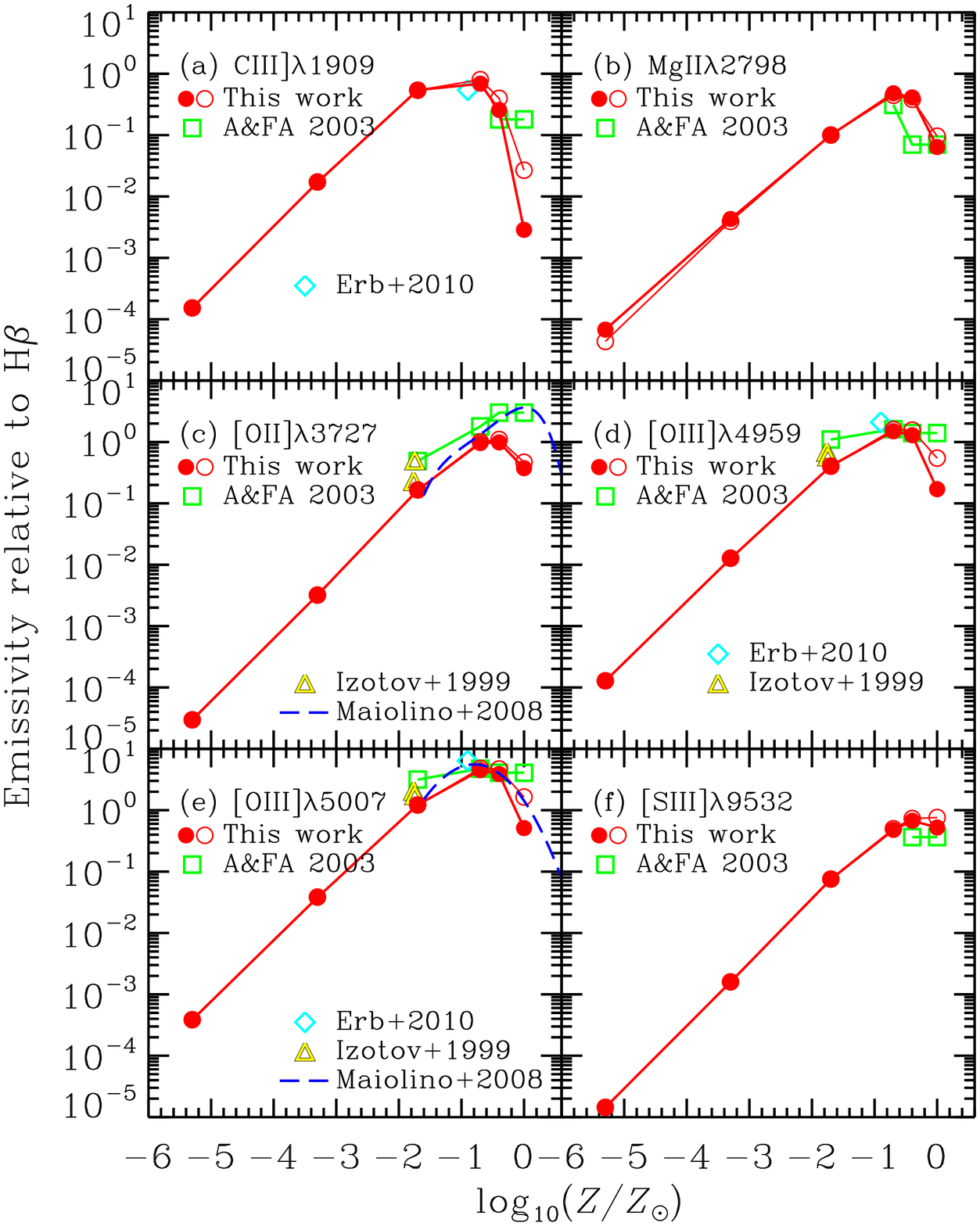}
}
\caption{{\em Left:} Predicted spectra (lines plus nebular and stellar continuum, solid) from photoionization models
at different metallicities. \heii\ and few other lines are identified. From \citet{2005ASSL..327..479P}.
{\em Right:} 
Predicted emissivities relative to \hb\ of salected strong metal lines as a function of metallicity from photoionization models (filled circles).
Other empirical data are shown by open symbols and the dashed line. From  \citet{2011MNRAS.415.2920I}.
%The open squares are empirical results by Anders & Fritze-v. Alvensleben (2003). The open triangles are the results of I Zw 18 (northwest and southeast %components) measured by Izotov et al. (1999). The dashed lines are empirical results by Nagao et al. (2006).
 }
\label{fig_lines}
\end{figure}

The relative line intensities of emission lines from all elements can be predicted from 
photoionization models, such as {\em Cloudy} \citep{Ferland98}.
Figure \ref{fig_lines} shows the predicted spectra and strengths of several of the strongest
lines from photoionization models as a function of metallicity. The left panel shows the overall 
SED including H and He lines for nebulae computed with blackbody spectra from calculations of
\citet{2005ASSL..327..479P}; the right panel the relative intensity of selected metal lines in the (rest frame) UV and
optical domain computed using realistic SEDs from synthesis models \citep{2011MNRAS.415.2920I}.
As well known, the intensity of the forbidden oxygen lines peaks at low metallicities ($Z/\zsun \sim$ 1/50),
and it is found to decrease monotonously to lower metallicity.
Measuring e.g.\ a line ratio of \Oiii/\hb $<0.1$ would indicate metallicities below $<10^{-3}$ \zsun\
(or alternatively $Z/\zsun$\ga 4), which could be feasible with NIRSpec on the James Webb Space Telescope 
(JWST), according to \citet{2005ASSL..327..479P}  and \citet{2011MNRAS.415.2920I}.
For more information on the feasibility of such observations see also the Chapter by Stiavelli.

%***mid-IR !?

% % % % % % % % % % % % % % % % % % % % % % % % % % 
\subsection{Peculiar colors}
Several recent papers have examined/proposed various criteria which could be used to distinguish
PopIII from other stellar populations at high redshift, based on photometry/colors 
\citep[see][]{2011MNRAS.415.2920I,2011MNRAS.418L.104Z,2011ApJ...740...13Z}.

As already discussed above (Sect.\ \ref{s_uvslope}), the UV slope is difficult to use as a metallicity indicator due to
degeneracies between age/star-formation history and metallicity, and due to the contribution of nebular continuum
emission which considerably ``dampens" the metallicity dependence of the UV slope on metallicity. In cases of a
large escape fraction of the Lyman continuum flux the latter difficulty is of course not present.
Predicted magnitudes and colors for different populations and variable escape fractions are presented in 
\citet{2011MNRAS.415.2920I,2011ApJ...740...13Z}.

\citet{2011ApJ...740...13Z} have proposed that PopIII galaxies at $z \sim 8$ with low escape fractions can be distinguished
from metal-rich objects based on two colors between 4.4 and 7.7 \micron, which could be measured with the
NIRCam and MIRI instrument on the the JWST. The basis for such a distinction is the presence of strong H lines (\ha) and 
the absence of \Oiiiab\ in these filters at these redshifts.

\begin{figure}[htb]
\sidecaption
\centering{
\includegraphics[width=0.6\textwidth]{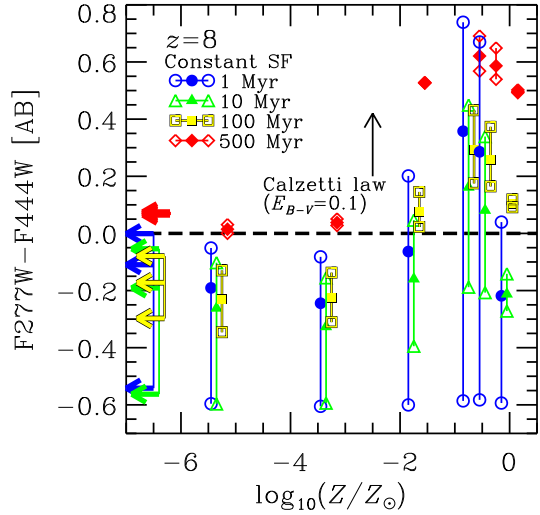}}
\caption{Predicted JWST/NIRCAM F277W--F444W colors of $z = 8$ galaxies as a function of metallicity. 
Different symbols/colors illustrate variations of the star formation history.
Vertical lines connect models with varying Lyman continuum escape fraction.
The horizontal dashed line is a proposed criterium to separate very metal-poor cases from higher metallicity cases.
From \citet{2011MNRAS.415.2920I}.}
\label{fig_color}
\end{figure}

Other criteria to select very metal-poor galaxies exploit the expectation of very blue colors between the 
UV and optical domain, which can to some extent be quite independent of nebular emission, i.e.\ of the escape fraction
\citep{2011MNRAS.415.2920I}.
Indeed, in case of strong nebular emission, the UV continuum should be strong with a Balmer jump
in emission, and emission lines weak in the optical (due to low metallicity). If in contrast the escape fraction was large, 
one recovers again the intrinsically blue stellar spectrum. In both cases one therefore expects a fairly blue
UV--optical color. At $z \sim 8$, for example, the color could be measured between 2.7 and 4.4 \micron\ with
the JWST; the corresponding predictions are shown in Fig.\ \ref{fig_color}. They show how such a blue color could
be exploited to select very metal-poor ($Z \la 10^{-2}$ \zsun) galaxies, although in some conditions an overlap 
also exists with higher metallicities.

PopIII or very metal-poor galaxies with strong \lya\ emission (cf.\ Sect.\ \ref{s_lya}) could also be detected
due to the excess caused by this line. For example at $z \sim$ 8--9 this can cause an
unusually blue $J-H$ color, as shown by \cite{2003RMxAC..16..225P} and discussed in detail by
\cite{2011MNRAS.418L.104Z}. In any case, it is clear that nebular emission (both lines and continua) 
can significantly contribute to broad-band fluxes \citep{SdB09,SdB10,2011arXiv1111.6373S}, and 
\citet{2011A&A...536A..72S} have recently shown that SED modeling techniques can recover the 
strength of \lya\ emission from current broad-band photometric surveys of $z \sim$ 3--6 galaxies. 
This demonstrates that various photometric criteria and SED fitting methods should also be able to
select peculiar objects, such as very metal-poor galaxies and PopIII dominated objects.

%%%%%%%%%%%%%%%%%%%%%%%%%%%%%%%%%%%%%%%%%%%%%%%%%%
\section{Conclusion}
\label{s_conclude}

As should be clear from the onset, evolutionary synthesis models are an important, fundamental tool
to interpret a wide variety of extra-galactic observations, from the nearby Universe to the most distant,
first galaxies. They are commonly used to render complex state-of-the-art simulations of the Universe
``visible", i.e.\ to translate physical properties of simulated galaxies into observables.
Finally, they are also key for many ``prospective'' studies, such as for the preparation of new missions,
to guide observers searching for Population III objects etc. 

Conceptually simple, synthesis models basically gather what is known about star-formation
(the IMF and star-formation history), stellar evolution and  atmospheres, and some additional
emission processes, to predict the temporal evolution of the spectrum of an integrated stellar population.
For this reason, synthesis models can basically only be as good/reliable as their input physics
is. Therefore regular updates are necessary, and it is important to keep problems
and limitations of these ``ingredients" and assumptions in mind.

For the spectral modeling of first galaxies, very metal-poor galaxies, and Population III objects, one of the major
unknowns is certainly the stellar initial mass function. The evolution of massive stars especially at low metallicities,
observationally inaccessible in the nearby Universe, remains also uncertain and efforts are ongoing to
properly account for the effects of rotation, magnetic fields and related processes. 
The star formation histories of young, distant galaxies are probably complex and fairly stochastic at least
on short time scales. Some of these questions have been addressed above, at least schematically 
(Sect.\ \ref{s_intro}). 

In any case, combining the best of our knowledge evolutionary synthesis models represent a very useful 
and important tool for studies of galaxy formation and evolution, from the nearby Universe back to first galaxies.

%%%%%%%%%%%%%%%%%%%%%%%%%%%%%%%%%%%%%%%%%%%%%%%%%%
\begin{acknowledgement}
I have benefitted from many interesting and stimulating discussions and collaborations with colleagues 
during the past years. I wish to thank them here collectively for this. 
%I also wish to thank the reader for his/her tolerance for non-exhaustive citations in many parts of this review.
I acknowledge support from the Swiss National Science Foundation.

\end{acknowledgement}
%%%%%%%%%%%%%%%%%%%%%%%%%%%%%%%%%%%%%%%%%%%%%%%%%%
\bibliographystyle{aa}
\bibliography{PopIII_all}

\end{document}